\documentclass[onecolumn]{autart}

\usepackage{tikz}
\usepackage{graphicx}
\usepackage{calc}
\usepackage{etoolbox}
\usepackage{xcolor}

\usepackage{graphicx}     
\let\theoremstyle\relax
\usepackage{amsmath, amssymb,bbm,amsthm,bbold}
\usepackage{enumitem}
\usepackage{comment}
\usepackage{algorithm}
\usepackage{wrapfig}

\usepackage{algpseudocode}
\usepackage{natbib}
\usepackage[english]{babel}
\usepackage{epstopdf}
\usepackage{tikz}
\usepackage{chngcntr}
\usepackage{apptools}
\let\theoremstyle\relax

\newtheorem{theorem}{Theorem}
\newtheorem{remark}[theorem]{Remark}
\newtheorem{lemma}[theorem]{Lemma}
\newtheorem{definition}[theorem]{Definition}
\newtheorem{proposition}[theorem]{Proposition}
\newtheorem{corollary}[theorem]{Corollary}
\newtheorem{assumption}{Assumption}

\newtheorem{example}{Example}

\DeclareMathOperator*{\rank}{rank}
\DeclareMathOperator*{\diag}{diag}

\DeclareMathOperator{\spn}{span}

\makeatletter
\def\@journal{Journal of the Franklin Institute}
\def\@date{August 31, 2026}
\makeatother

\begin{document}
\begin{frontmatter}
\title{On the Discernibility of Topology Changes in Networked LTI Systems}
\author[unipd]{Giulio Fattore}\ead{fattoregiu@dei.unipd.it}
\quad\text{and}\quad
\author[unipd]{Maria Elena Valcher}\ead{meme@dei.unipd.it}  

\address[unipd]{Department of Information Engineering, University of Padova, Via Gradenigo 6/B, 35131 Padova, Italy} 


\begin{abstract} 
This   article investigates the discernibility of topology changes in networked linear time-invariant systems, with particular focus on edge disconnection. We introduce a new notion of discernibility motivated by a limitation of the definition commonly adopted in the literature. In consensus networks, topology changes are inherently indiscernible for all initial conditions belonging to the synchronization subspace.
To overcome this issue, we first introduce the notion of {\em dynamical discernibility of a topology change} for elementary consensus networks. This notion explicitly disregards the intrinsic indiscernibility associated with states in the consensus subspace. We characterize this property for directed communication graphs and then specialize the resulting conditions to topology changes arising from either the disconnection of an edge or a variation in its weight. We then extend the analysis to networks of homogeneous agents by introducing {\em discernibility modulo synchronization}, which factors out the unavoidable indiscernibility associated with synchronized trajectories. We derive a general characterization showing how discernibility depends jointly on the topology change and on the dynamics of the individual agents.
Finally, we specialize the analysis to undirected communication graphs and show that, under mild additional assumptions, discernibility modulo synchronization is equivalent to the dynamical discernibility of the corresponding topology change. This establishes a direct connection between the two notions and extends existing results on the discernibility of topology variations in dynamical networks. The resulting characterizations are then further specialized to the case of edge disconnection.
\end{abstract}

\end{frontmatter}

\section{Introduction}

Driven by the increasing size and complexity of cyber-physical systems, over the past decades there has been a progressive shift from centralized solutions toward distributed implementations. Distributed approaches have found widespread application in areas such as multi-agent robotics \cite{Rizk_ACM2019}, state estimation \cite{DUIO_TAC,Han2019,Nuno3}, output regulation \cite{Su_TAC2012}, and optimization \cite{Molzahn_TSG2017,Nedic_AR2018,YANG_ARC2019}. Distributed architectures offer several advantages, including improved scalability, enhanced fault tolerance, and greater responsiveness to real-time changes. At the same time, however, they introduce new challenges related to coordination, communication, and robustness against disruptions.
In addition, in many distributed problems, the structure of the underlying communication network plays a fundamental role in ensuring convergence to the desired solution. Consequently, changes in the network topology caused, for instance, by unknown faults affecting the communication links \cite{CHEN_AUT2020}, denial-of-service (DoS) attacks \cite{Amin_DOS2009,Dolk_TAC2017,pasqualettiTAC2013}, or the disconnection of nodes and edges \cite{ICSTCC2019}, may compromise the conditions required for convergence and ultimately prevent the system from reaching the desired solution.
For these reasons, the prompt detection of changes in the network topology has become crucial for preserving the correct operation of distributed systems and preventing potentially irreversible performance degradation. 

Various approaches have been proposed in the literature to address this broad class of problems. 
In \cite{Rahimian_TCNS2015}, the problem of detecting and isolating link failures in networks of linear continuous-time dynamical systems is considered. The proposed methodology relies on the analysis of discontinuities arising in suitable derivatives of the node outputs following a change in the network topology.
In \cite{RahimianACC2012,RahimianCDC2012}, the problem of detecting single or multiple link failures in networks of single-integrator agents is investigated. The authors introduce the notion of a distinguishable flow graph and derive sufficient conditions for failure detectability in terms of internodal distances.

A different perspective is developed in \cite{Battistelli_CDC2015}, where integrator networks subject to topology variations are modeled as multi-mode systems, with the switching signal determining the currently active network configuration. Within this framework, link-failure detection and isolation can be addressed by studying the distinguishability of different modes, namely by determining whether distinct network configurations can generate identical system trajectories. This problem was previously addressed in the literature on multimode and switched systems under the notions of discernibility and mode observability \cite{Baglietto_AUT2007,Baglietto_AUT2014,Battistelli_AUT2013,Vidal_CDC2002}.

All the aforementioned contributions focus on networks of integrators and therefore exploit the specific structure of their dynamics. Relaxing this assumption makes it possible to study topology variations in a substantially broader class of networked dynamical systems, thereby extending the scope and applicability of the resulting detection and discernibility conditions.

To the best of our knowledge, one of the first fundamental contributions in this direction is \cite{BattistelliTesi} (see also \cite{patil2019indiscernible,Roy_TCNS2020}), where the discernibility of topology variations is investigated for networks of dynamical systems interconnected through an undirected graph, under both complete and partial access to the state measurements. These works introduce the notions of an \emph{indiscernible state} and an $M$-\emph{indiscernible state}, where $M$ denotes the output matrix of the overall multi-agent system. In particular, topology variations caused by edge or node disconnections are considered, and conditions are derived under which such variations can be detected from the network dynamics. Under suitable assumptions on the spectra of the matrices describing the agent dynamics and on the spectrum of the graph Laplacian, necessary and sufficient conditions for state discernibility are established. In particular, these results show that, under the aforementioned spectral assumptions (see \cite{Roy_TCNS2020}), state discernibility is equivalent to the discernibility of the corresponding topology variation. A notion of \emph{output discernibility}, distinct from $M$-indiscernibility, is also introduced and characterized.

Building on these results, \cite{ECC2019,ICSTCC2019} investigate the effects of edge and node disconnections on consensus in simple networks modeled as discrete-time integrators. It is shown that the disconnection of an edge in an undirected communication network does not alter the final consensus value, provided that the network remains connected. In contrast, the disconnection of a node may change the consensus value even when connectivity of the remaining network is preserved. The discernibility of these faults is then analyzed under both full and partial state measurements, and a procedure is proposed to recover the original consensus value following a node disconnection.

In \cite{Hao_TAC2023}, the analysis is extended to directed networks of homogeneous linear time-invariant (LTI) agents. The authors first consider a general network model in which the interconnection topology is represented by a generic network matrix, without imposing a Laplacian structure. Necessary and sufficient conditions for discernibility are derived in terms of generalized eigenvectors of matrices determined jointly by the agent dynamics and the eigenvalues of the network matrix. The analysis is subsequently specialized to relevant classes of topology variations, including edge and node disconnections, as well as to the case of a diagonalizable network matrix.  The authors then focus on Laplacian-based multi-agent systems over undirected graphs and establish necessary and sufficient conditions relating topology discernibility to both the agent dynamics and the spectral properties of the graph Laplacian. 
In addition, the subspace of initial conditions for which the nominal and modified network configurations are indiscernible is explicitly characterized.

The problem of discernibility under partial measurements is further addressed in \cite{Hao_AUT2024}, where the same authors study topology variations when the agent states are not directly accessible and only a subset of output measurements is available. Necessary and sufficient conditions for $M$-discernibility (as defined in \cite{BattistelliTesi}) are derived for a generic networked system. These conditions are formulated in terms of the intersection between eigenspaces associated with the system matrices before and after the topology variation and the kernel of the overall output matrix. Furthermore, by exploiting the Jordan canonical transformation and, in particular, generalized Jordan chains,
the authors obtain additional conditions characterizing output-indiscernible topology variations.

A related problem is considered in \cite{Fan_TAC2024} for directed communication graphs and networks of homogeneous scalar systems under partial state measurements. The authors adopt a notion of output discernibility of topology variations that differs from the one introduced in \cite{BattistelliTesi}. They first characterize the set of output-indiscernible states and derive necessary and sufficient conditions for a given initial state to be output-discernible. These conditions are expressed in terms of the Jordan chains associated with the eigenvalues of the system matrices corresponding to the nominal and modified topologies. Then, a characterization of output discernibility directly in terms of the two system matrices, again through their Jordan-chain structure, is derived. An algorithm for computing the set of output-indiscernible states is also proposed. Finally, edge and node disconnections in networks governed by consensus dynamics are examined through a numerical example.

Topology variations have also been studied in the more general framework of heterogeneous descriptor systems. In \cite{KUSTERS_AUT2017}, networks of heterogeneous single-input single-output (SISO) Differential-Algebraic Equations (DAEs) are considered. Necessary and sufficient conditions for the indiscernibility of a topology change are derived, with the notable property that they can be verified using only the eigenvalues and eigenvectors of the nominal network configuration. 
However, the work does not provide a characterization of the set of initial network states for which the topology variation remains undetectable.
This limitation is addressed in \cite{patil2019indiscernible}, which considers the more general setting of networks of heterogeneous multi-input multi-output (MIMO) DAEs interconnected through diffusive coupling. The authors first characterize the subspace of initial conditions giving rise to indiscernible nominal and modified network configurations. Based on this characterization, they derive necessary and sufficient conditions for indiscernibility in terms of common eigenspaces of the two configurations. They further show that the indiscernibility property can be assessed solely from the eigenspaces of the nominal network configuration.

The paper \cite{Wang_TeF2025} investigates the discernibility of topology variations in heterogeneous multi-agent systems  with directed communication graphs. It shows that node heterogeneity can fundamentally affect discernibility, so that topology changes discernible in homogeneous networks may become indiscernible, and vice versa. The authors derive necessary and sufficient conditions for state-based discernibility, together with specialized characterizations for link variations and node disconnections. Necessary and sufficient conditions for \(M\)-discernibility are also established.
\smallskip

In most of  the previous works, the adopted notion of discernibility seems to be not adequate to provide a meaningful 
characterization of the performance of a consensus network in the presence of topology changes. Indeed, when a topology variation occurs in a network, the corresponding Laplacian matrix is reconfigured,  resulting in a different Laplacian matrix. Nevertheless, the consensus subspace is invariant under any graph Laplacian and, consequently, a consensus state produces the same state evolution for the original system and the one with  modified topology. Consequently, when dealing with a network of homogeneous agents interacting through a consensus mechanism,  every synchronization state is indiscernible with respect to the considered topology variation. This implies that the existing notion of discernibility can never be satisfied for such networked systems.

To overcome this limitation, in this article we introduce a  notion of discernibility specifically tailored to the detection of topology changes in the Laplacian matrix. It aligns with the viewpoint first adopted in 
\cite{ECC2019,ICSTCC2019} and
excludes the intrinsic indiscernibility associated with the consensus subspace.  We refer to this property as \emph{dynamical discernibility of a topology change}. We first characterize dynamical discernibility for directed graphs and then specialize the resulting conditions to the cases of edge disconnections and undirected graphs. 

We subsequently consider networks of homogeneous agents. Since homogeneous multi-agent systems admit synchronization trajectories that are inherently indiscernible under the standard notion of discernibility, we eliminate this trivial source of ambiguity by introducing the notion of \emph{discernibility modulo synchronization} of a topology change. We first derive  general characterizations of this property for directed networks and then adapt them to the case of undirected networks. Since the resulting characterizations require either the computation of the unobservable subspace or the evaluation of the PBH test for a large matrix pair, both of which may be computationally demanding, we identify mild additional assumptions under which discernibility modulo synchronization for undirected networks can be assessed through considerably simpler conditions.
In particular, when the coupling matrix $B$, which weights the disagreement between neighboring agents, is nonsingular, we show that dynamical discernibility of a topology change is both necessary and sufficient for discernibility modulo synchronization. When  $B$ is singular, additional assumptions on the spectra of certain matrices associated with the agent descriptions and the Laplacian eigenvalues lead  to simpler characterizations of discernibility. Indeed, under such assumptions, dynamical discernibility is again shown to be necessary and sufficient for discernibility modulo synchronization. Finally, we specialize our analysis to the case of edge disconnection and derive necessary and sufficient conditions for discernibility modulo synchronization.

The remainder of the paper is organized as follows. In Section~\ref{secLaplacian}, we investigate the effects of topology changes on elementary consensus models associated with directed communication graphs. In particular, Subsection \ref{subsec.sign.edge.disc}  
focuses on the case of an edge disconnection (or of a change of weight  of a single edge).
  In Section~\ref{secMAS}, we investigate the discernibility of topology variations in homogeneous multi-agent systems (MASs) with directed communication graphs. Section~\ref{secMASundir} specializes these results to undirected communication graphs and identifies mild additional assumptions under which the characterization of discernibility modulo synchronization simplifies considerably. These assumptions include the case of a nonsingular coupling matrix $B$ (Subsection~\ref{subsec.nssb}) and suitable spectral conditions involving the agent dynamics and the eigenvalues of the Laplacian matrix (Subsection~\ref{subsec.spectralcond}). Finally, Subsection~\ref{subsec.MASedgedisc} specializes the analysis to edge disconnections. Section~\ref{sec.conclusions} concludes the paper and discusses directions for future research.
\medskip

{\bf Notation.} The sets of real numbers, nonnegative real numbers and nonnegative integers are denoted by $\mathbb{R},   {\mathbb R}_+$ and $\mathbb{Z}_+$, respectively. Given two integers $h,k$, with $h \le k$, we let $[h,k]$ denote the set $\{h,h+1,\dots,k\}$.  
The symbol  $I_n$ denotes the identity matrix of size $n$, $\mathbb{0}_{m\times n}$ the zero matrix of size $m\times n$, $\mathbb{0}_{n}$ and ${\mathbb 1}_n$ are the all-zero and all-one vectors of size $n$, respectively. Suffixes will be omitted when the dimensions are irrelevant or can be deduced from the context. The $i$-th vector of the canonical basis in ${\mathbb R}^n$ (where $n$ is always clear from the context) is denoted by $\mathbb{e}_i.$
The $(i,j)$-th entry of a matrix $M \in \mathbb R^{m\times n}$ is 
  denoted   by $[M]_{ij}$ or by $m_{ij}$,
  while a matrix whose entries are $m_{ij}$ is denoted by $M = [m_{ij}]$. 
We use $\ker(M)$ to represent the {\em kernel space} of $M$, and ${\rm im}(M)$ to represent its {\em column space} (or {\em image}). 
The {\em spectrum} of a square matrix $M$ is denoted by $\sigma(M)$ and is the set of all its eigenvalues.
An {\em eigenpair}  of $M$ is a pair $(\lambda, v)$, where $\lambda$ is an eigenvalue of $M$ and $v\ne {\mathbb 0}$ an associated eigenvector.
A square symmetric matrix $M$ is {\em positive semidefinite} if  $x^\top M x \ge 0$ for every $x\ne {\mathbb 0}$.
The {\em Kronecker product} is denoted by $\otimes$.
Given a set ${\mathcal S}$, the symbol $|{\mathcal S}|$ denotes its cardinality. Given a family of vectors $\{v_i\}_{i=1}^p, v_i\in {\mathbb R}^n$, the symbol ${\spn}(v_1, \dots, v_p)$ represents the vector space generated by the vectors $v_1, \dots, v_p$.

A {\em directed graph} is a triple $\mathcal G = (\mathcal V, \mathcal E, {\mathcal W})$, where ${\mathcal V}=[1,N]$ is the set of nodes, ${\mathcal E} \subseteq {\mathcal V}\times {\mathcal V}$ is the set of directed edges and ${\mathcal W}\in {\mathbb R}_+^{N\times N}$ is the weight (adjacency) matrix. The (directed) edge $(j,i)$, from $j$ to $i$, belongs to ${\mathcal E}$ if and only if $[{\mathcal W}]_{ij} > 0$. 
A \emph{directed path} from node $i$ to node $j$ is a sequence of directed edges
$(i,i_1), (i_1,i_2), \dots, (i_k,j)$ each belonging to ${\mathcal E}$. 
A {\em directed spanning tree} of ${{\mathcal G}}$ is a subgraph of ${{\mathcal G}}$ containing all the nodes in ${\mathcal V}$ and satisfying the following properties: \emph{i)} it contains no closed path, and \emph{ii)} it has a unique root node from which every other node is reachable through a directed path.
 The {\em Laplacian}  associated with ${\mathcal G} = ({\mathcal V}, {\mathcal E}, {\mathcal W})$ is the matrix  ${\mathcal L}\in {\mathbb R}^{N \times N}$ whose entries are defined as follows
\begin{align*}
[{\mathcal L}]_{ij} :=  
\begin{cases}
- [{\mathcal W}]_{ij}, & i \ne j ,\\
\sum_{k = 1}^{N}{[{\mathcal W}]_{ik}}, & i = j.
\end{cases}
\end{align*}
An {\em undirected graph} is a graph  ${\mathcal G} = ({\mathcal V}, {\mathcal E}, {\mathcal W})$ whose weight matrix  ${\mathcal W}\in \mathbb{R}_+^{N\times N}$ is  symmetric. When so, the Laplacian ${\mathcal L}$ is symmetric and positive semidefinite. 
An undirected graph that contains a spanning tree is {\em connected}.

\section{The effect of graph topology changes in a consensus model}\label{secLaplacian}
The first topic we are interested in exploring is the effect of a topology change on an elementary consensus model, represented in the continuous-time case by the state-space model
\begin{equation}
\dot{x}(t) = - {\mathcal L} x(t),
\qquad t\in {\mathbb R}_+,
\label{CT_consensus}
\end{equation}
and in the discrete-time case by
\begin{equation}
x(k+1) = (I_N - \alpha {\mathcal L}) x(k),
\qquad k\in {\mathbb Z}_+,
\label{DT_consensus}
\end{equation}
where ${\mathcal L}\in {\mathbb R}^{N \times N}$ is the Laplacian matrix associated with the directed communication graph ${\mathcal G}=({\mathcal V}, {\mathcal E}, {\mathcal W})$,   ${\mathcal V}=[1,N]$, while $\alpha$ is any positive real scalar such that $I_N-\alpha {\mathcal L}$ is Schur stable, namely,
\[
0 < \alpha < \frac{1}{\max_i [{\mathcal L}]_{ii}}.
\]
Suppose now that the communication network ${\mathcal G}$ undergoes a topology change, yielding the graph $\bar {\mathcal G}=({\mathcal V}, \bar {\mathcal E}, \bar {\mathcal W})$, with associated Laplacian matrix $\bar {\mathcal L}$. We will denote such topology change with the symbol ${\mathcal L} \to \bar {\mathcal L}$ \cite{Fan_TAC2024,Hao_TAC2023,Hao_AUT2024}.
 In the continuous-time case the resulting system is
\begin{equation}
\dot{x}(t) = - \bar {\mathcal L} x(t),
\qquad t\in {\mathbb R}_+,
\label{CT_consensus_mod}
\end{equation}
while in the discrete-time case  is
\begin{equation}
x(k+1) = (I_N - \alpha \bar {\mathcal L}) x(k),
\qquad k\in {\mathbb Z}_+.
\label{DT_consensus_mod}
\end{equation}
\begin{remark} \label{rem.consornot}
While \eqref{CT_consensus_mod}
is always a consensus model, this is not necessarily true for \eqref{DT_consensus_mod}, since the state matrix $I_N-\alpha \bar {\mathcal L}$ of the modified model
may fail to remain Schur stable. Nevertheless, there are classes of topology changes that preserve this property.
This is the case, for instance, when the topology change consists of removing an edge, say $(j,i)\in {\mathcal E}\setminus \bar {\mathcal E}$. In this case, the new Laplacian matrix $\bar {\mathcal L}$ is related to the original one ${\mathcal L}$ by
\begin{equation}
\bar {\mathcal L}
=
{\mathcal L}
-
[{\mathcal W}]_{ij}\,
{\mathbb e}_i({\mathbb e}_i-{\mathbb e}_j)^{\top},
\label{newLaplcian_ED}
\end{equation}
and, since
\[
\max_i[\bar{\mathcal L}]_{ii}
\le
\max_i[{\mathcal L}]_{ii},
\]
the matrix $I_N-\alpha\bar{\mathcal L}$ remains Schur stable.  Similar considerations apply when the weight of an existing edge is reduced, provided that it remains positive. This property, however, does not generally hold when a new edge is added or the weight of an existing edge is increased, as these operations may increase the largest diagonal entry of the Laplacian. This issue, however, is not relevant to the subsequent analysis.
\end{remark}

We begin by introducing the notion of state indiscernibility separately for the continuous-time and discrete-time models. We then show that the two definitions can be unified into a common framework.

\begin{definition}[State Indiscernibility - First Version] \cite{BattistelliTesi}
Consider a continuous-time consensus model \eqref{CT_consensus} or a discrete-time consensus model \eqref{DT_consensus},  associated with the Laplacian matrix
     $\mathcal{L}\in\mathbb{R}^{N\times N}$, and let  $\mathcal{\bar L}\in\mathbb{R}^{N\times N}$ be a distinct Laplacian matrix.  We say that $x_0\in\mathbb{R}^N$  
     is {\em indiscernible with respect to the topology change}
$\mathcal{L}\rightarrow\bar{\mathcal L}$ if, starting from the same initial condition $x_0$, the original and modified consensus models, namely, systems \eqref{CT_consensus} and \eqref{CT_consensus_mod} in continuous time, or systems \eqref{DT_consensus} and \eqref{DT_consensus_mod} in discrete time,  generate identical state trajectories.
\end{definition}

It is easy to see that in the continuous-time case, 
$x_0\in\mathbb{R}^N$ is indiscernible with respect to the topology change $\mathcal{L}\rightarrow\bar{\mathcal L}$ if
\begin{align*}e^{-{\mathcal L}t}x_0= e^{-\bar {\mathcal L}t}x_0, \qquad \forall\ t\in {\mathbb R}_+,\end{align*}
and by making use of the power series expansion of the exponential of a matrix, the previous condition holds if and only if 
\begin{align}\label{eq.indiscenibility}
        \mathcal{L}^kx_0=\mathcal{\bar L}^kx_0,\quad \forall k\in\mathbb{Z}_+.
    \end{align}
Similarly, if we consider the discrete-time model \eqref{DT_consensus} and the one resulting from the topology change \eqref{DT_consensus_mod}, a state
    $x_0\in\mathbb{R}^N$ is indiscernible with respect to the topology change $\mathcal{L}\rightarrow\bar{\mathcal L}$ if
\begin{align}\label{eq.indiscenibility.2}
(I_N-\alpha \mathcal{L})^kx_0=(I_N-\alpha\mathcal{\bar L})^kx_0,\quad \forall k\in\mathbb{Z}_+,
\end{align}
which is easily seen to be equivalent, in turn, to condition \eqref{eq.indiscenibility}.
Therefore, in the following, we will resort to the following definition of indiscernible state for a topology change.

\begin{definition}[State Indiscernibility - Second Version]
    Let $\mathcal{L}\in\mathbb{R}^{N\times N}$ and $\mathcal{\bar L}\in\mathbb{R}^{N\times N}$ be two distinct Laplacian matrices. We say that the state $x_0\in\mathbb{R}^N$ 
    is {\em indiscernible for the topology change} $\mathcal{L}\rightarrow\mathcal{\bar L}$ if 
    condition \eqref{eq.indiscenibility} holds.
\end{definition}

In the rest of the paper, we will steadily refer to the latter definition.

\begin{remark}
    It is readily verified that any consensus state, namely any state of the form
$c\mathbb{1}_N$, with $c\in\mathbb{R}$, is a (static) equilibrium of both the
continuous-time systems \eqref{CT_consensus} and \eqref{CT_consensus_mod}, as well as of the discrete-time systems \eqref{DT_consensus} and \eqref{DT_consensus_mod}. Consequently,
\[
\mathcal{L}^k(c\mathbb{1}_N)
=
\bar{\mathcal L}^{\,k}(c\mathbb{1}_N),
\qquad
\forall\,k\in\mathbb{Z}_+.
\]
This implies that every consensus state is indiscernible with respect to any topology change
$\mathcal{L}\rightarrow\bar{\mathcal L}$.
This observation motivates the following notion of discernibility  of a topology change in which consensus states are regarded as the only admissible indiscernible states (see \cite{ECC2019,ICSTCC2019}).
\end{remark}

\begin{definition}[Dynamically discernible topology change]\label{def.dyn.dis.tc}
The topology change $\mathcal{L}\rightarrow\mathcal{\bar L}$ is {\em dynamically discernible} if the only indiscernible states belong to $\spn(\mathbb{1}_N)$.
\end{definition}

We will provide a complete characterization of the previous property under this basic assumption (that is typically introduced when dealing with consensus models as \eqref{CT_consensus} and  \eqref{DT_consensus}).

 \begin{assumption}\label{ass.sp.tree}
    The directed graph ${\mathcal G}=({\mathcal V}, {\mathcal E}, {\mathcal W})$ contains a spanning tree.
\end{assumption}
Note that by \cite[Theorem 6.6]{FB-LNS} this is equivalent to imposing that $0$ is a simple eigenvalue of the Laplacian matrix $\mathcal{L}$.
 
The following proposition provides a family of characterizations of the dynamic discernibility of a topology change.

\begin{proposition}\label{prop.top.cg.1}
Let Assumption \ref{ass.sp.tree} hold. The following facts are equivalent:
\begin{itemize}
\item[i)] The topology change $\mathcal{L}\rightarrow\bar{\mathcal{L}}$ is dynamically discernible.
\item[ii)]
The Laplacian matrices $\mathcal{L}$ and $\bar{\mathcal{L}}$ have no common eigenpair $(\lambda,v)$ with $\lambda\neq 0$.
    \item [iii)] There is no  $\lambda\ne 0$  such that
    \begin{align*}
        {\rank}
\begin{bmatrix}
    \lambda I-\mathcal{L} & \\
    & \lambda I-\bar{\mathcal{L}}\\
    \hline
    I & -I
\end{bmatrix}
<2N.
    \end{align*}
     \item[iv)] The pair $\left(
    \begin{bmatrix}
        \mathcal{L} & \\
        & \bar{\mathcal{L}}
    \end{bmatrix},
    \begin{bmatrix}
        I_N & -I_N
    \end{bmatrix}
    \right)$ is reconstructable.
     \item [v)] There is no  $\lambda\ne 0$  such that
    \begin{align*}
        {\rank}
\begin{bmatrix}
    \lambda I-\mathcal{L} \\
    \hline
   \mathcal{L} -\bar{\mathcal{L}}
\end{bmatrix}
<N.
\end{align*}
    \item[vi)] The pair $(\mathcal{L},\mathcal{L}-\mathcal{\bar L})$ is reconstructable.
     \item[vii)] 
$X^{no}(\mathcal{L},\mathcal{L}-\mathcal{\bar L})=\spn(\mathbb{1}_N)$.
\item[viii)] ${\rm ker} ({\mathcal L}^N) \supseteq X^{no}(\mathcal{L},\mathcal{L}-\mathcal{\bar L}).$
\end{itemize}
\end{proposition}
\begin{proof}
$i) \Rightarrow ii)$ \ Suppose, by contradiction, that there exists a common eigenpair $(\lambda,v)$, with $\lambda\ne0$  (and $v\ne {\mathbb 0}$), then for every $k\ge 0$
    $$
        \mathcal{L}^kv=\lambda^kv=\mathcal{\bar L}^k v.
    $$
    On the other hand, since $\lambda\ne 0$, then  $v$ cannot belong to ${\spn}(\mathbb{1}_N)$. Therefore, $v$ is an indiscernible non-consensus state for the topology change $\mathcal{L}\rightarrow\bar{\mathcal{L}}$, and such change is not dynamically discernible.\\
    
$ii) \Rightarrow iii)$\ Suppose, by contradiction,  that 
there exists $\lambda\ne 0$   such that
    \begin{align*}
        {\rank}
\begin{bmatrix}
    \lambda I-\mathcal{L} & \\
    & \lambda I-\bar{\mathcal{L}}\\
    \hline
    I & -I
\end{bmatrix}
<2N.
    \end{align*}
    This implies that there exists $v \ne {\mathbb 0}$ such that
\[\begin{bmatrix}
    \lambda I-\mathcal{L} & \\
    & \lambda I-\bar{\mathcal{L}}\\
    \hline
    I & -I
\end{bmatrix}\begin{bmatrix}v \cr v\end{bmatrix} = {\mathbb 0}.\]
Consequently,
${\mathcal L}v = \lambda v = \bar {\mathcal L}v,$
and this contradicts $ii)$.\\ 

$iii) \Rightarrow v)$\ Suppose, by contradiction, that there exists $\lambda \ne 0$ such that 
\begin{align*}
        {\rank}
\begin{bmatrix}
    \lambda I-\mathcal{L} \\
    \hline
   \mathcal{L} -\bar{\mathcal{L}}
\end{bmatrix}
<N.
\end{align*}
This implies that there exists $v \ne {\mathbb 0}$ such that 
\begin{align*}
    \begin{bmatrix}
    \lambda I-\mathcal{L} \\
    \hline
   \mathcal{L} -\bar{\mathcal{L}}
\end{bmatrix} v = {\mathbb 0}.
\end{align*}
But then it is immediate to see that 
\begin{align*}
\begin{bmatrix}
    \lambda I-\mathcal{L} & \\
    & \lambda I-\bar{\mathcal{L}}\\
    \hline
    I & -I
\end{bmatrix}\begin{bmatrix} v \cr v\end{bmatrix} = {\mathbb 0},
\end{align*}
thus contradicting  $iii)$.\\

$v) \Rightarrow vii)$\ If $v)$ holds, this means that $\lambda =0$ is the only eigenvalue of the unobservable subsystem of the pair $({\mathcal L}, {\mathcal L}-\bar {\mathcal L})$. On the other hand, by Assumption \ref{ass.sp.tree}, $0$ is a simple eigenvalue of ${\mathcal L}$ and this means that the matrix of the unobservable subsystem has dimension $1$. But this also means that the dimension of the unobservable subsystem is $1$ and since it is always true that $\spn({\mathbb 1}_N) \subseteq X^{no}(\mathcal{L},\mathcal{L}-\mathcal{\bar L})$, it follows that
$\spn({\mathbb 1}_N) = X^{no}(\mathcal{L},\mathcal{L}-\mathcal{\bar L})$.\\

$vii) \Rightarrow i)$\ 
Assume that the topology change ${\mathcal L} \to \bar {\mathcal L}$ is not dynamically discernible, and hence there exists $v\not\in \spn({\mathbb 1}_N)$ such that
$\mathcal{L}^kv=\mathcal{\bar L}^kv$ for every $k\ge0$.
It is easy to see that the previous 
condition implies
$({\mathcal L} - \bar {\mathcal L}) {\mathcal L}^k v = {\mathbb 0}$ for every $k\in {\mathbb Z}_+$, and hence
there exists 
$v\not\in \spn({\mathbb 1}_N)$ belonging to 
$X^{no}(\mathcal{L},\mathcal{L}-\mathcal{\bar L})$.
\\

$iii) \Leftrightarrow iv)$,  $v) \Leftrightarrow vi)$ and
$vi) \Leftrightarrow viii)$
\ Follow from standard Systems Theory results.
\end{proof}

\begin{remark} The discernibility of topology variations has been extensively investigated in the literature
\cite{BattistelliTesi,Fan_TAC2024,Hao_TAC2023,Hao_AUT2024,patil2019indiscernible}. The first characterizations were derived for general LTI systems, without explicitly exploiting the structural properties of networked systems. As a consequence, several conditions in Proposition~\ref{prop.top.cg.1} are closely related to previously established discernibility criteria. In particular, conditions $ii)$ and $iii)$ are closely related to results in \cite{BattistelliTesi}, while conditions $ii)$ and $vi)$ bear similarities to those in \cite{Hao_TAC2023}. However, even when specialized to networked systems, these results lead to different characterizations because they do not exclude the case in which the initial state is a consensus state, namely,
$x_0=c\mathbb{1}_N.$
In contrast, the present work explicitly rules out this trivial source of indiscernibility through the notion of \emph{dynamical discernibility of topology changes} introduced in Definition~\ref{def.dyn.dis.tc}. Consequently, although some of the conditions in Proposition~\ref{prop.top.cg.1} are structurally related to existing discernibility criteria, they play a fundamentally different role by providing the theoretical foundation for our notion of dynamical discernibility in networked systems.
\end{remark}

To conclude the section, we want to provide another characterization of the dynamic discernibility of the topology change ${\mathcal L} \to \bar {\mathcal L}$ that will be useful for the following general analysis.

To this end, it is convenient to introduce the following notation. 
Let us assume that $\mathcal{L}$ has $r$ distinct eigenvalues, i.e., $\sigma(\mathcal{L})=\{\lambda_1,\lambda_2,\dots,\lambda_r\}$, with  $\lambda_1:=0$ and $\lambda_i \ne \lambda_j$ for $i\ne j$. Let $g_i$ and $a_i$ denote the geometric and algebraic multiplicities, respectively, of the eigenvalue $\lambda_\ell, \ell\in[1,r]$. 
We denote by $T\in\mathbb{C}^{N\times N}$ a nonsingular matrix that transforms the Laplacian matrix $\mathcal{L}$ into its Jordan canonical form, namely,
\begin{align}
\label{Jordan1}
    J:=T^{-1}\mathcal{L} T=\begin{bmatrix}
        0&&&\\
        &J_2&&\\
        &&\ddots&\\
        &&&J_r
    \end{bmatrix} \in {\mathbb C}^{N\times N}, 
    \end{align}
    where 
    \begin{align}
    \label{Jordan2}
    J_\ell &:=\begin{bmatrix}
        J_{\ell,1}&&\\
        &\ddots&\\
        &&J_{\ell,g_\ell}
    \end{bmatrix}\in\mathbb{C}^{a_\ell\times a_\ell} \quad {\rm and}\quad
    \quad J_{\ell,h}:= \begin{bmatrix} \lambda_\ell & 1 & 0 &\dots & 0\cr   
0 & \lambda_\ell & 1 & \dots &0\cr
\vdots &\vdots &\ddots &\ddots& \vdots \cr
0 & 0 &0 & \ddots & 1\cr
0 & 0 &0 & \dots & \lambda_\ell\end{bmatrix}\in\mathbb{C}^{d_{\ell,h}\times d_{\ell,h}}.
\end{align}
By making use of the transformation matrix $T$, we also define
\begin{equation}
\label{defW}
W := (\mathcal{L}-\bar{\mathcal L})T \in {\mathbb C}^{N\times N}.
\end{equation}
  The matrix \(W\) is first partitioned according to the Jordan block structure of \(J\). Each block \(W_\ell\) is then further partitioned conformably with the corresponding Jordan miniblock structure.That is,
\begin{align} 
   & W :=\begin{bmatrix}
        w_1&W_2&\dots&W_r
    \end{bmatrix}, \label{eq.W1}\\
    {\rm for}\ \ell\in [2,r], \quad  & W_\ell:=\begin{bmatrix}
        W_{\ell,1}&\cdots&W_{\ell,g_\ell}
    \end{bmatrix},\quad W_{\ell,h}:=\begin{bmatrix}
        w_{\ell,h}^{(1)} &\star&\cdots&\star
    \end{bmatrix}, \label{eq.W2}
\end{align}
where we denoted by $w_{\ell,h}^{(1)}$ the first column of the block $W_{\ell,h}$ and by $\star$ the subsequent columns (whose value will not be relevant).
Note that, by Assumption \ref{ass.sp.tree}, $0$ is a simple eigenvalue of ${\mathcal L}$, and hence the corresponding block in $W$ consists of a single column that we have denoted by the symbol $w_1$. On the other hand, the first column of $T$ is an eigenvector of ${\mathcal L}$ corresponding to the zero eigenvalue and hence it must necessarily be of the form $c \cdot {\mathbb 1}_N$. This implies that 
${\mathcal L} T {\mathbb e}_1= {\mathbb 0}$ as well as  $\bar {\mathcal L} T {\mathbb e}_1= {\mathbb 0}$, and hence  $w_1 = ({\mathcal L}-\bar {\mathcal L})T {\mathbb e}_1= {\mathbb 0}$. 

\begin{proposition} \label{prop.sette}
Let Assumption \ref{ass.sp.tree} hold. Suppose that 
$T$ is a nonsingular transformation matrix that reduces ${\mathcal L}$ to Jordan form, and assume that $J=T^{-1}{\mathcal L}T$ and $W = ({\mathcal L}-\bar {\mathcal L})T$ are described as  in \eqref{Jordan1}-\eqref{Jordan2} and \eqref{eq.W1}-\eqref{eq.W2}, respectively. Finally, set
\begin{equation}
    W_\ell^{(1)} := \begin{bmatrix} w_{\ell,1}^{(1)}  & w_{\ell,2}^{(1)} & \dots & w_{\ell,g_\ell}^{(1)}\end{bmatrix}.
    \label{eq.Welle1}
\end{equation}
The topology change $\mathcal{L}\rightarrow\bar{\mathcal{L}}$ is dynamically discernible if and only if, for every   $\ell\in[2,r]$, the matrix $W_\ell^{(1)}$ has full column rank.
\end{proposition}
\begin{proof}
By Proposition \ref{prop.top.cg.1}, the topology change $\mathcal{L}\rightarrow\bar{\mathcal{L}}$ is dynamically discernible if and only if   
    \begin{align*}
        {\rank}
\begin{bmatrix}
    \lambda I-\mathcal{L} \\
    \hline
   \mathcal{L} -\bar{\mathcal{L}}
\end{bmatrix}
=N, \qquad \forall\ \lambda \ne 0.
\end{align*} 
 Taking advantage of the matrix $T$ and the fact that the distinct eigenvalues of ${\mathcal L}$ are $\{0,\lambda_2, \dots, \lambda_r\}$, the previous condition can be equivalently rewritten as
 \begin{equation}
 \label{rankcondJW}
        {\rank}
\begin{bmatrix}
    \lambda_\ell I- J \\
    \hline
   W
\end{bmatrix}
=N, \qquad \forall \ \ell\in [2,r].
\end{equation} 
By applying the PBH observability test to the pair $(J, W)$, where $J$ is in Jordan canonical form, the previous condition holds if and only if for every $\ell\in [2,r]$ the family of vectors $\{w_{\ell,h}^{(1)}\}_{h\in[1,g_\ell]}$ is linearly independent,  which amounts to saying that $W_\ell^{(1)}$ has full column rank.
\end{proof}

\begin{example}
    Consider a continuous time system
    \begin{align*}
        \dot x(t)= - \mathcal{L}x(t),\quad \text{where}\quad \mathcal{L}=\begin{bmatrix}
1 & -1 & 0 & 0 & 0 & 0 \\
-1.5 & 2.5 & -1 & 0 & 0 & 0 \\
-1 & -2 & 3 & 0 & 0 & 0 \\
-0.2 & -1 & -0.3 & 1.5 & 0 & 0 \\
-1 & 0 & 0 & 0 & 1 & 0 \\
-1 & 0 & 0 & 0 & 0 & 1
\end{bmatrix}.
    \end{align*}
    The Laplacian matrix $\mathcal{L}$ corresponds to the directed graph $\mathcal{G}$ (see the graph on the left in Figure \ref{fig:topology-weight-change}).
    Suppose that a topology variation occurs due to the disconnection of the edge $(2,1)$ and the change in weight of the edges  $(2,3)$  and $(1,4)$. The resulting system is described by
     \begin{align*}
        \dot {\bar x}(t)=- \mathcal{\bar L}x(t),\quad \text{where}\quad \mathcal{\bar L}=\begin{bmatrix}
0 & 0 & 0 & 0 & 0 & 0 \\
-1.5 & 2.5 & -1 & 0 & 0 & 0 \\
-1 & -2.2 & 3.2 & 0 & 0 & 0 \\
-1.2 & -1 & -0.3 & 2.5 & 0 & 0 \\
-1 & 0 & 0 & 0 & 1 & 0 \\
-1 & 0 & 0 & 0 & 0 & 1
\end{bmatrix}.
    \end{align*}

  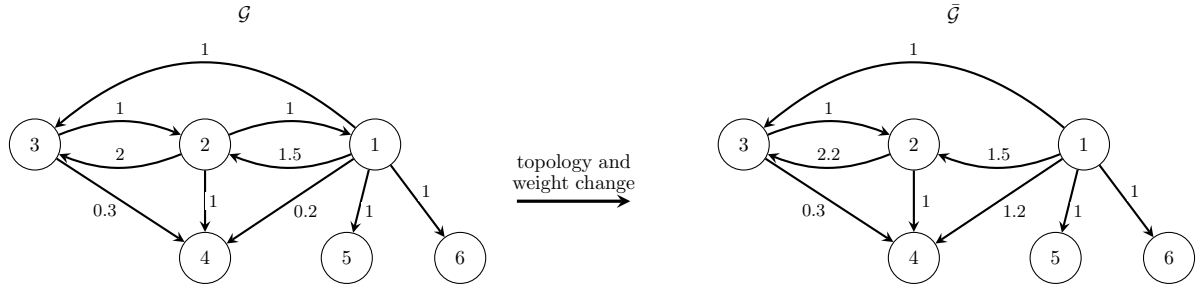
\begin{figure}
\begin{center}
\begin{tikzpicture}[
 scale=0.75,
    transform shape,
    node/.style={
        circle,
        draw,
        minimum size=9mm,
        inner sep=0pt
    },
    edge/.style={
        ->,
        thick
    },
    weight/.style={
        draw=none,
        fill=white,
        inner sep=1.5pt,
        font=\small
    },
    >=stealth
]


\node[node] (1) at (6,2) {$1$};
\node[node] (2) at (3,2) {$2$};
\node[node] (3) at (0,2) {$3$};

\node[node] (4) at (3,0) {$4$};
\node[node] (5) at (5.5,0) {$5$};
\node[node] (6) at (7.5,0) {$6$};

\draw[edge]
    (1) to[bend left=22]
    node[weight, midway, above,yshift=2pt] {$1.5$}
    (2);

\draw[edge]
    (2) to[bend left=22]
    node[weight, midway, above,yshift=2pt] {$1$}
    (1);

\draw[edge]
    (3) to[bend left=22]
    node[weight, midway, above,yshift=2pt] {$1$}
    (2);

\draw[edge]
    (2) to[bend left=22]
    node[weight, midway, above,yshift=2pt] {$2$}
    (3);

\draw[edge]
    (1) to[out=140,in=40,looseness=1.15]
    node[weight, midway, above,yshift=2pt] {$1$}
    (3);

\draw[edge]
    (1) --
    node[weight, midway, below right] {$0.2$}
    (4);

\draw[edge]
    (2) --
    node[weight, midway, right] {$1$}
    (4);

\draw[edge]
    (3) to[ right=12]
    node[weight, midway, below left] {$0.3$}
    (4);

\draw[edge]
    (1) --
    node[weight, midway, below right] {$1$}
    (5);

\draw[edge]
    (1) --
    node[weight, midway, above right] {$1$}
    (6);

\node[draw=none] at (3.7,4.3) {$\mathcal{G}$};


\draw[->, very thick]
    (8.5,1) -- (10.5,1)
    node[midway, above, align=center, draw=none]
    {topology and\\ weight change};


\begin{scope}[xshift=12.5cm]

\node[node] (b1) at (6,2) {$1$};
\node[node] (b2) at (3,2) {$2$};
\node[node] (b3) at (0,2) {$3$};

\node[node] (b4) at (3,0) {$4$};
\node[node] (b5) at (5.5,0) {$5$};
\node[node] (b6) at (7.5,0) {$6$};

\draw[edge]
    (b1) to[bend left=22]
    node[weight, midway, above,yshift=2pt] {$1.5$}
    (b2);


\draw[edge]
    (b3) to[bend left=22]
    node[weight, midway, above,yshift=2pt] {$1$}
    (b2);

\draw[edge]
    (b2) to[bend left=22]
    node[weight, midway, above,yshift=2pt] {$2.2$}
    (b3);

\draw[edge]
    (b1) to[out=140,in=40,looseness=1.15]
    node[weight, midway, above,yshift=2pt] {$1$}
    (b3);

\draw[edge]
    (b1) --
    node[weight, midway, below right] {$1.2$}
    (b4);

\draw[edge]
    (b2) --
    node[weight, midway, right,xshift=2pt] {$1$}
    (b4);

\draw[edge]
    (b3) to[ right=12]
    node[weight, midway, below left] {$0.3$}
    (b4);

\draw[edge]
    (b1) --
    node[weight, midway,below  right] {$1$}
    (b5);

\draw[edge]
    (b1) --
    node[weight, midway,above right] {$1$}
    (b6);

\node[draw=none] at (3.7,4.3) {$\bar{\mathcal{G}}$};

\end{scope}

\end{tikzpicture}
\end{center}

\caption{Topology and edge-weight change from $\mathcal{G}$ to
$\bar{\mathcal{G}}$.}
\label{fig:topology-weight-change}
\end{figure}
We want to verify wether the topology change occurred is dynamically discernible. To do so we make use of Proposition \ref{prop.sette}. Let $T$ be the transformation matrix that brings $\mathcal{L}$ in Jordan form, namely 
\begin{align*}
J=T^{-1}\mathcal{L}T =
\begin{bmatrix}
0 & 0 & 0 & 0 & 0 & 0 \\
0 & 1.5 & 0 & 0 & 0 & 0 \\
0 & 0 & 2.2192 & 0 & 0 & 0 \\
0 & 0 & 0 & 4.2808 & 0 & 0 \\
0 & 0 & 0 & 0 & 1 & 0 \\
0 & 0 & 0 & 0 & 0 & 1
\end{bmatrix},\quad  \text{where}\quad T =
\begin{bmatrix}
1 & 0 & -1.2192 & -3.2808 & 0 & 0 \\
1 & 0 & 1.4865 & 10.7635 & 0 & 0 \\
1 & 0 & 2.2462 & -14.2462 & 0 & 0 \\
1 & 1 & -2.6647 & -2.0978 & 0 & 0 \\
1 & 0 & 1 & 1 & 1 & 0 \\
1 & 0 & 1 & 1 & 0 & 1
\end{bmatrix},
\end{align*}
and compute the matrix 
\begin{align*}
W =(\mathcal{L}-\mathcal{\bar L})T=
\begin{bmatrix}
0 & 0 & -2.7057 & -14.0443 & 0 & 0 \\
0 & 0 & 0 & 0 & 0 & 0 \\
0 & 0 & -0.1519 & 5.0019 & 0 & 0 \\
0 & -1.0000 & 1.4455 & -1.1830 & 0 & 0 \\
0 & 0 & 0 & 0 & 0 & 0 \\
0 & 0 & 0 & 0 & 0 & 0
\end{bmatrix}.
\end{align*}
It is readily verified that the columns of \(W\) corresponding to the Laplacian eigenvalue \(\lambda=1\) are identically zero. Hence the topology change is dynamically indiscernible.
\end{example}
\medskip

 \subsection{Topology changes in an undirected (and connected) graph}
 \smallskip
 
 In the special case of an undirected graph $\mathcal{G}$, the Laplacian matrix ${\mathcal L}$ is symmetric, and the characterization provided by Proposition~\ref{prop.sette} admits a considerably simpler form.  
 Since this result will play a key role in the sequel, we derive it explicitly below.

  We observe that if ${\mathcal L}$ is symmetric, then it is diagonalizable, and hence the Jordan form \eqref{Jordan1} holds for 
\[
J_\ell= \lambda_\ell I_{a_\ell},
\quad
\ell\in[2,r].
\]
Similarly,   the matrix $({\mathcal L}-\bar{\mathcal L})T$ can be partitioned as in \eqref{eq.W1}, where each block $W_\ell$ is a  real matrix  of size $N \times a_\ell$.  Note that in this case $W_\ell^{(1)}=W_\ell$.

\begin{corollary}\label{corotto}
If ${\mathcal G}$ is an undirected and connected graph (namely Assumption \ref{ass.sp.tree} holds), whose corresponding Laplacian ${\mathcal L}$ has $r$ distinct eigenvalues 
$\lambda_1=0<\lambda_2<\cdots<\lambda_r,$ 
then the topology change $\mathcal{L}\rightarrow\bar{\mathcal{L}}$ is dynamically discernible if and only if each matrix $W_\ell$, $\ell\in [2,r]$, 
  has full column rank.
\end{corollary}

\subsection{Single edge disconnection or change of weight}\label{subsec.sign.edge.disc}
\smallskip

In this subsection, we consider topology changes consisting of a modification of the weight of a single edge. This includes, as a special case, assigning zero weight to an edge, which corresponds to its removal.
Without loss of generality, let $(j,i)\in\mathcal E$ denote the affected edge.
Then (see \eqref{newLaplcian_ED}) the difference between the original and the modified Laplacian matrices can be written   \cite{BattistelliTesi} as
\begin{align}
\mathcal{L}-\bar{\mathcal L}
=
\alpha {\mathbb e}_i(\mathbb e_i-\mathbb e_j)^\top,
\end{align}
where $\alpha \ne 0$ denotes the corresponding weight variation.
In this case, by exploiting the special structure of the difference matrix $\mathcal{L}-\bar{\mathcal L}$, conditions $v), vi)$, $vii)$ and $viii)$  of Proposition \ref{prop.top.cg.1} can be equivalently rewritten in simplified form as follows:
\begin{itemize}
    \item [$v)$] There is no  $\lambda\ne 0$  such that
    \begin{align*}
        {\rank}
\begin{bmatrix}
    \lambda I-\mathcal{L} \\
    \hline
   (\mathbb e_i-\mathbb e_j)^\top
\end{bmatrix}
<N.
\end{align*}
    \item[$vi)$] The pair $(\mathcal{L},(\mathbb e_i-\mathbb e_j)^\top)$ is reconstructable.
     \item[$vii$)] 
$X^{no}(\mathcal{L},(\mathbb e_i-\mathbb e_j)^\top)=\spn(\mathbb{1}_N)$.
\item[$viii$)] ${\rm ker} ({\mathcal L}^N) \supseteq X^{no}(\mathcal{L},(\mathbb e_i-\mathbb e_j)^\top).$
\end{itemize}
A more informative characterization can be obtained by resorting to the pair $(J,W)$, where
$J=T^{-1}{\mathcal L}T$ is the Jordan canonical form of ${\mathcal L}$, and
$W$ is defined in \eqref{defW}.

\begin{proposition} \label{ED_eigenvectorentries}
Let Assumption~\ref{ass.sp.tree} hold. The topology change
$\mathcal{L}\rightarrow\bar{\mathcal{L}}$ resulting from a variation in the weight of the edge
$(j,i)\in{\mathcal E}$ is dynamically discernible if and only if the following conditions are satisfied:
\begin{enumerate} 
\item[i)] ${\mathcal L}$ is cyclic.
\item[ii)] if $v_\ell$, $\ell\in[2,r]$, denotes an eigenvector of ${\mathcal L}$ associated with the eigenvalue $\lambda_\ell$, then
\begin{equation} \label{viNOTvj}
[v_\ell]_i\neq [v_\ell]_j,
\qquad \forall\,\ell\in[2,r].
\end{equation}
\end{enumerate}
\end{proposition}
\begin{proof}
Suppose, without loss of generality, that the Jordan canonical form of ${\mathcal L}$,
$J=T^{-1}{\mathcal L}T$, is described by \eqref{Jordan1} and \eqref{Jordan2}. The transformation matrix $T$ can be partitioned as
\[
T=
\begin{bmatrix}
T_1&T_2&\cdots&T_r
\end{bmatrix},
\]
where $T_\ell$, $\ell\in[1,r]$, denotes the block associated with the eigenvalue $\lambda_\ell$. In turn, each block admits the partition
\[
T_\ell=
\begin{bmatrix}
T_{\ell,1}&T_{\ell,2}&\cdots&T_{\ell,g_\ell}
\end{bmatrix},
\]
conformably with the decomposition of $J_\ell$ into Jordan miniblocks. The first column of each subblock, namely $T_{\ell,h}\mathbb e_1$, is an eigenvector of ${\mathcal L}$ associated with the eigenvalue $\lambda_\ell$.

By Proposition~\ref{prop.top.cg.1}, the topology change
$\mathcal L\rightarrow\bar{\mathcal L}$ induced by a variation in the weight of the edge $(j,i)\in{\mathcal E}$ is dynamically discernible if and only if
\[
\rank
\begin{bmatrix}
\lambda I-\mathcal L\\
\hline
(\mathbb e_i-\mathbb e_j)^\top
\end{bmatrix}
=N,
\qquad
\forall\,\lambda\neq0,
\]
or, equivalently,
\[
\rank
\begin{bmatrix}
\lambda_\ell I-J\\
\hline
(\mathbb e_i-\mathbb e_j)^\top T
\end{bmatrix}
=N,
\qquad
\forall\,\ell\in[2,r].
\]
Arguing as in the proof of Proposition~\ref{prop.sette}, the above rank condition holds if and only if the following two properties are satisfied:
\begin{enumerate}
\item[\emph{a)}]
Each eigenvalue $\lambda_\ell$, $\ell\in[2,r]$, is associated with a single Jordan block. Since $\lambda_1=0$ is simple, this is equivalent to requiring that ${\mathcal L}$ be cyclic, i.e., $g_\ell=1$ for every $\ell\in[1,r]$. Consequently, $T_\ell=T_{\ell,1}$ for every $\ell\in[2,r]$.

\item[\emph{b)}]
For every $\ell\in[2,r]$,
\[
(\mathbb e_i-\mathbb e_j)^\top T_\ell\mathbb e_1
=
(\mathbb e_i-\mathbb e_j)^\top v_\ell
\neq\mathbb 0,
\]
which is precisely condition~\eqref{viNOTvj}.
\end{enumerate}
\hfill
\end{proof}

\begin{remark}  \label{remark10}
    When the communication graph ${\mathcal G}$ is undirected and connected, the disconnection (or weight variation) of the edge $(j,i)$ is modeled by
\[
{\mathcal L}-\bar{\mathcal L}
=
\alpha({\mathbb e}_i-{\mathbb e}_j)
({\mathbb e}_i-{\mathbb e}_j)^\top,
\]
for some $\alpha \ne 0$. Conditions $v)$-$viii)$ of Proposition~\ref{prop.top.cg.1} coincide with those derived at the beginning of this same subsection for directed graphs. On the other hand, the symmetry of ${\mathcal L}$ simplifies condition $i)$ of Proposition~\ref{ED_eigenvectorentries}, which  reduces to the fact all the   eigenvalues of ${\mathcal L}$ must be distinct (i.e., $r=N$).
\end{remark}
 
\begin{example}
Consider the continuous-time system
\begin{align*}
    \dot{x}(t) = -\mathcal{L}x(t),
\quad
\text{where}
\quad 
\mathcal{L} =
\begin{bmatrix}
1 & -1 & 0 & 0 \\
-1 & 2 & 0 & -1 \\
0 & 0 & 1 & -1 \\
0 & -1 & -1 & 2
\end{bmatrix}.
\end{align*}
The Laplacian matrix ${\mathcal L}$ corresponds to the (undirected) graph ${\mathcal G}$  (see the graph on the left in Figure \ref{fig:topology-change.ex.1}).
\medskip

Suppose that a topology change occurs due to the disconnection of the edge $(2,4)$. 
The resulting graph $\bar{\mathcal{G}}$ (see the graph on the right in Figure \ref{fig:topology-change.ex.1}) is no longer connected and consists of two connected components, namely $\{1,2\}$ and $\{3,4\}$, and
the modified system is described by
\begin{align*}
    \dot{\bar{x}}(t) = -\mathcal{\bar{L}}\bar{x}(t),
\quad
\text{where}
\quad 
\mathcal{\bar{L}} =
\begin{bmatrix}
1 & -1 & 0 & 0 \\
-1 & 1 & 0 & 0 \\
0 & 0 & 1 & -1 \\
0 & 0 & -1 & 1
\end{bmatrix}.
\end{align*}

\begin{figure}
\begin{center}
\begin{tikzpicture}[
    node/.style={
        circle,
        draw,
        minimum size=9mm,
        inner sep=0pt
    },
    edge/.style={
        thick
    },
    >=stealth
]


\node[node] (1) at (0,2) {$1$};
\node[node] (2) at (3,2) {$2$};
\node[node] (3) at (0,0) {$3$};
\node[node] (4) at (3,0) {$4$};

\draw[edge] (1) -- (2);
\draw[edge] (2) -- (4);
\draw[edge] (3) -- (4);

\node at (1.5,3) {$\mathcal{G}$};


\draw[->, very thick]
    (4.2,1) -- (6.2,1)
    node[midway, above, align=center]
    {edge $(2,4)$\\ disconnection};


\node[node] (b1) at (7.5,2) {$1$};
\node[node] (b2) at (10.5,2) {$2$};
\node[node] (b3) at (7.5,0) {$3$};
\node[node] (b4) at (10.5,0) {$4$};

\draw[edge] (b1) -- (b2);
\draw[edge] (b3) -- (b4);

\node at (9,3) {$\bar{\mathcal{G}}$};

\end{tikzpicture}\end{center}

\caption{Topology change due to the disconnection of edge $(2,4)$.}
\label{fig:topology-change.ex.1}
\end{figure}
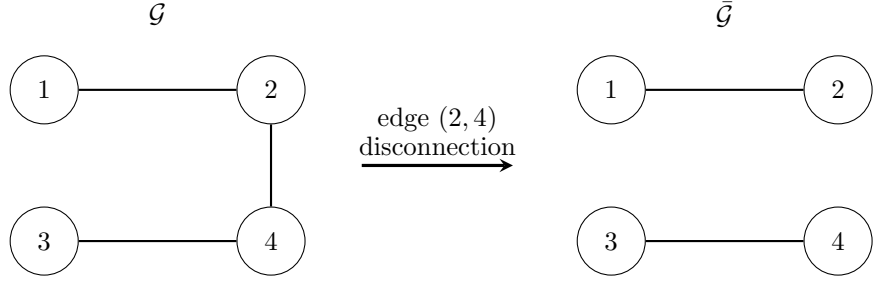
Since the topology change considered in this example corresponds to an edge disconnection   in an undirected graph, we investigate its dynamical discernibility by means of Proposition~\ref{ED_eigenvectorentries}  and the subsequent remark.

The  spectrum of $\mathcal{L}$ is given by $\sigma(L)=\{\lambda_1, \lambda_2, \lambda_3, \lambda_4\}= \{0,\,0.5858,\,2,\,3.4142\}$. Since all the eigenvalues of $\mathcal{L}$ are distinct, $\mathcal{L}$ is cyclic, hence condition $i)$  of Proposition~\ref{ED_eigenvectorentries} is satisfied. 
The eigenvectors of ${\mathcal L}$ corresponding to such an ordered quadruple of eigenvalues appear as columns of  the transformation matrix
\begin{align*}
T =
\begin{bmatrix}
0.5 & 0.6533 & -0.5 & 0.2706 \\
0.5 & 0.2706 & {\color{red}0.5} & -0.6533 \\
0.5 & -0.6533 & -0.5 & -0.2706 \\
0.5 & -0.2706 & {\color{red}0.5} & 0.6533
\end{bmatrix} = \begin{bmatrix} v_1 & v_2 & v_3 & v_4\end{bmatrix}.
\end{align*}
We now consider the second condition of Proposition~\ref{ED_eigenvectorentries}.  For the disconnection of edge $(2,4)$, the entries corresponding to nodes $2$ and $4$ of the eigenvectors associated with the non-zero eigenvalues must be considered. 
It is easy to see that $[v_3]_2=[v_3]_4=0.5$. It follows that condition $ii)$ of Proposition~\ref{ED_eigenvectorentries} is not satisfied, and hence the topology change $\mathcal{L}\rightarrow\bar{\mathcal{L}}$ is dynamically indiscernible.
Indeed,  let us initialize both systems with the   initial condition
\begin{align*}
    x(0)=\bar{x}(0)=x_0=
\begin{bmatrix}
0.5\\
1.5\\
0.5\\
1.5
\end{bmatrix}.
\end{align*}
Figure~\ref{fig:indiscernibility-trajectories.ex2.2} shows that the state trajectories of the two systems starting from $x_0$ are identical and, hence, indiscernible.
\begin{figure}
    \centering
    \includegraphics[width=0.47\linewidth]{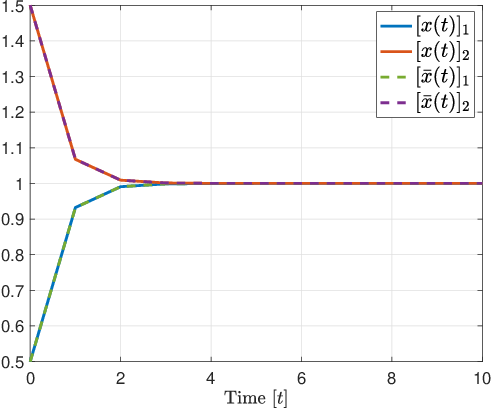}
    \includegraphics[width=0.47\linewidth]{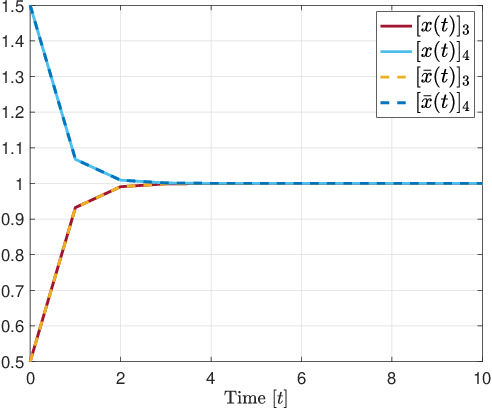}
    \caption{State evolution of the original system and of the system obtained after disconnecting edge $(2,4)$. On the left, evolution of the first and second state components of the two systems; on the right, evolution of the third and fourth state components.} 
    \label{fig:indiscernibility-trajectories.ex2.2}
\end{figure}
Let us now suppose that   both systems are initialized with the  initial condition
\begin{align*}
    x(0)=\bar{x}(0)=x_0=
\begin{bmatrix}
0\\
2\\
0.5\\
1.5
\end{bmatrix}.
\end{align*}
For the original connected system, the asymptotic consensus value is given by the average of the initial states, namely
\begin{align*}
    x_{\mathrm{av}}=\frac{1}{4}\mathbb{1}^{\top}x_0
    =
    \frac{0+2+0.5+1.5}{4}=1.
\end{align*}
Therefore,
    $\lim_{t\rightarrow\infty}x(t)
    =\mathbb{1}_4$.
After the edge disconnection, the two connected components evolve independently. The consensus value associated with the first connected component $\{1,2\}$ is
\begin{align*}
    \bar{x}_{\mathrm{av},1}=
    \frac{x_1(0)+x_2(0)}{2}=
    \frac{0+2}{2}=1,
\end{align*}
whereas the consensus value associated with the second connected component $\{3,4\}$ is
\begin{align*}
    \bar{x}_{\mathrm{av},2}=
    \frac{x_3(0)+x_4(0)}{2}=
    \frac{0.5+1.5}{2}=1.
\end{align*}
Consequently, it is also true that
    $\lim_{t\rightarrow\infty}\bar{x}(t)
    =\mathbb{1}_4.$
 This means that
for this choice of the initial condition, the original connected system and the system obtained after the edge disconnection asymptotically converge to the same  consensus value. Therefore, the occurrence of the topology change cannot be inferred by considering the steady-state behavior alone. Nevertheless, the two systems exhibit different transient responses, as shown in Fig.~\ref{fig:discernibility-trajectories}, and hence the the state $x_0$ is dynamically discernible.  

\begin{figure}
    \centering
    \includegraphics[width=0.47\linewidth]{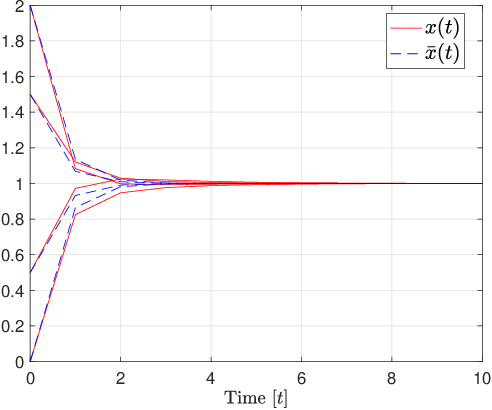}
    \includegraphics[width=0.47\linewidth]{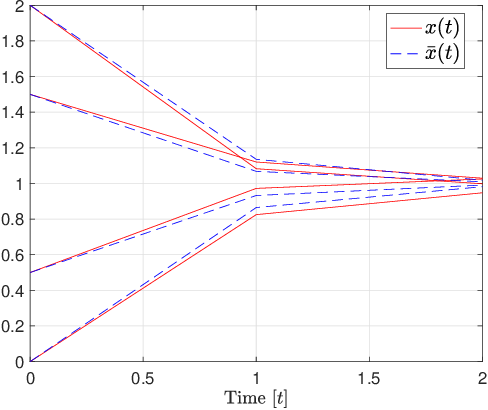}
    \caption{State evolution of the original system and of the system obtained after the disconnection of edge $(2,4)$.   On the left the evolution in the time interval $[0,10]$; on the right its restriction to $[0,2]$. Although the two state trajectories, $x(t)$ and $\bar x(t)$, converge to the same asymptotic value, their transient responses are different. }
    \label{fig:discernibility-trajectories}
\end{figure}
\end{example}

\section{Detecting topology variations in  MASs}\label{secMAS}

In the remainder of the paper, we focus on discrete-time homogeneous multi-agent systems (MASs) described by higher-order dynamics rather than by simple integrator models.

Consider a discrete-time system comprised of $N$ homogeneous agents communicating through a directed graph ${\mathcal G}=({\mathcal V}, {\mathcal E}, {\mathcal W})$, with ${\mathcal V}=[1,N]$, satisfying Assumption \ref{ass.sp.tree}. 
 For every $i\in [1,N]$, the    dynamics of agent $i$ is described by
\begin{align*}
    x_i(t+1)=Ax_i(t)+B\sum_{j=1}^N w_{ij}(x_j(t)-x_i(t)), 
\end{align*}
 where $x_i(t)\in {\mathbb R}^n$ is the state of the $i$-th agent,  and $w_{ij}$ is the $(i,j)$-th entry of the weight matrix ${\mathcal W}$, representing the weight that agent $i$ gives to the mismatch of its state from the one of agent $j$. $A$ and $B$ are $n\times n$ real matrices. We will refer to $B$, which weights the diffusive terms $x_i(t)-x_j(t)$, as the {\em coupling matrix} \cite{LuChen2005,ZHOU2008996}.
Let $\mathcal{L}$ be the 
Laplacian matrix associated with the graph $\mathcal{G}$. Then the overall system dynamics is described by
\begin{align}\label{eq.s1}
    x(t+1)=(I_N\otimes A -\mathcal{L}\otimes B)x(t).
\end{align}
Under the effect of  the topology change $\mathcal{L}\rightarrow\bar{\mathcal{L}}$, the multi-agent system becomes 
\begin{align}\label{eq.s2}
    \bar x(t+1)=(I_N\otimes A -\mathcal{\bar L}\otimes B) \bar x(t).
\end{align}
As in Section~\ref{secLaplacian}, our goal is to characterize the conditions under which a topology change can be detected from the observed state evolution.
We first observe that if, at some time $t\in{\mathbb Z}_+$, the network is in a {\em synchronized state}, namely,
\[
x_i(t)=x_j(t), \qquad \forall\, i,j\in[1,N],
\]
or, equivalently,
\[
x(t)=\mathbb{1}_N\otimes z(t),
\]
for some $z(t)\in\mathbb{R}^n$, then it remains synchronized at all subsequent times. Indeed,  using either \eqref{eq.s1} or \eqref{eq.s2}, it follows that
\[
x(t)=\mathbb{1}_N\otimes z(t)
\quad\Longrightarrow\quad
x(t+1)
=
\mathbb{1}_N\otimes (Az(t))
=: \mathbb{1}_N\otimes z(t+1),
\]
which shows that synchronization is preserved irrespective of the communication weights, and hence of the Laplacian matrix.
Consequently, the state trajectories generated by any initial condition of the form
$x_0=\mathbb{1}_N\otimes z_0$, $z_0\in {\mathbb R}^n$,
are identical for systems \eqref{eq.s1} and \eqref{eq.s2}. This motivates the following definitions.

\begin{definition}[Discernible modulo synchronization  topology change] \label{def.indiscMAS}
Consider the two multi-agent systems \eqref{eq.s1} and \eqref{eq.s2}, the latter obtained from the former as a result of the topology change $\mathcal{L}\rightarrow\bar{\mathcal{L}}$. A state $x_0\in {\mathbb R}^{nN}$ is {\em indiscernible} for the two MASs   \eqref{eq.s1} and \eqref{eq.s2} if
\begin{equation}
(I_N\otimes A -\mathcal{L}\otimes B)^k x_0 = 
(I_N\otimes A -\bar{\mathcal L}\otimes B)^k x_0, \qquad \forall\ k\in {\mathbb Z}_+.
\end{equation}
The two multi-agent systems \eqref{eq.s1} and \eqref{eq.s2} are {\em discernible modulo synchronization} if the only indiscernible states  are the synchronization   ones, i.e.,  $x_0=\mathbb{1}_n\otimes z_0$.
\end{definition}

We first establish a necessary condition for the discernibility modulo synchronization of the MASs \eqref{eq.s1} and \eqref{eq.s2}. This condition was first proved    in    \cite[Corollary~4]{Hao_TAC2023} for undirected networks. However, since we adopt a slightly different notion of indiscernibility, and for the sake of completeness, we provide an independent proof. As will become apparent in the sequel, this result plays an important role in the derivation of the subsequent characterizations.

\begin{lemma}\label{ABobs}
  Let Assumption \ref{ass.sp.tree} hold.  If the systems \eqref{eq.s1} and \eqref{eq.s2} are discernible modulo synchronization, then the pair $(A,B)$ is observable.
\end{lemma}
\begin{proof}
    If the pair $(A,B)$ is not observable, then there exists an eigenpair $(\lambda,z)$, with $\lambda\in\mathbb{C}$ and  $z\ne\mathbb{0}$, such that 
    \begin{align*}
        \begin{cases}
            Az=\lambda z\\
            Bz=\mathbb{0}.
        \end{cases}
    \end{align*}
  It follows that for every $w\in {\mathbb R}^N$ we have 
  \begin{align*}
       (I_N\otimes A-\mathcal{L}\otimes B)(w\otimes z)&=w\otimes Az - {\mathcal L}w \otimes Bz =w\otimes\lambda z = \lambda(w\otimes z)\\
       (I_N\otimes A-\bar{\mathcal L}\otimes B)(w\otimes z)&=w\otimes Az - \bar {\mathcal L}w \otimes Bz =w\otimes\lambda z = \lambda(w\otimes z),
  \end{align*}
  from which it follows that 
$$(I_N\otimes A-\mathcal{L}\otimes B)^k(w\otimes z)=
\lambda^k(w\otimes z) = (I_N\otimes A-\bar{\mathcal L}\otimes B)^k(w\otimes z),\quad  \forall\ k \in {\mathbb Z}_+.$$
  Hence $w\otimes z$ is indiscernible for \eqref{eq.s1} and \eqref{eq.s2} even when $w\notin\spn(\mathbb{1}_N)$.
\end{proof}
Lemma~\ref{LLdyndisc}, below, establishes another necessary condition for discernibility modulo synchronization. Although the relationship between the two notions of discernibility has already appeared in the literature (see \cite{BattistelliTesi,Hao_TAC2023}), this condition has never been explicitly identified as a necessary condition, since it has always been embedded within more comprehensive characterizations involving additional assumptions. Moreover, the existing results have been established only for undirected communication graphs and, again, by referring to a notion of discernibility differently from the one adopted in this paper.

\begin{lemma}\label{LLdyndisc}
     Let Assumption \ref{ass.sp.tree} hold.  If the systems \eqref{eq.s1} and \eqref{eq.s2} are discernible modulo synchronization, then the topology change $\mathcal{L}\rightarrow\bar{\mathcal{L}}$  is dynamically discernible.
\end{lemma}
\begin{proof}
Suppose that the topology change $\mathcal{L}\rightarrow\bar{\mathcal{L}}$ is not dynamically discernible. Hence, see $ii)$ of Proposition \ref{prop.top.cg.1}, there exists $(\lambda, v)$, with $\lambda \ne 0$ and $v \ne {\mathbb 0}$, such that 
$$\left\{\begin{array}{rcl}
{\mathcal L} v &=& \lambda v, \cr
\bar {\mathcal L} v &=& \lambda v.
\end{array}\right.
$$
Clearly $v\not\in \spn({\mathbb 1}_N)$.
Let $z\ne {\mathbb 0}$ be any eigenvector of $A-\lambda B$ corresponding to some eigenvalue $\mu$.
Then
\begin{align*} 
       (I_N\otimes A-\mathcal{L}\otimes B)(v\otimes z)&=v\otimes Az - {\mathcal L}v \otimes Bz  = v\otimes Az - \lambda v\otimes \ Bz = v\otimes (A-\lambda B)z \\
       &= v\otimes (\mu z) = \mu (v\otimes z) \\
       (I_N\otimes A-\bar{\mathcal L}\otimes B)(v\otimes z)&=v\otimes Az - \bar {\mathcal L}v \otimes Bz  = v\otimes Az - \lambda v\otimes \ Bz = v\otimes (A-\lambda B)z \\
       &= v\otimes (\mu z) = \mu (v\otimes z),
  \end{align*}
  from which it follows that 
$$(I_N\otimes A-\mathcal{L}\otimes B)^k(v\otimes z)=
\mu^k(v\otimes z) = (I_N\otimes A-\bar{\mathcal L}\otimes B)^k(v\otimes z), \quad\forall\ k \in {\mathbb Z}_+.$$
  Hence $v\otimes z$ is indiscernible for \eqref{eq.s1} and \eqref{eq.s2} for some $v\notin\spn(\mathbb{1}_N)$.
\end{proof}

\begin{remark}
    An alternative proof of this result could be obtained by resorting to the Jordan form of ${\mathcal L}$, $J=T^{-1}{\mathcal L}T$, and  to the matrix
    $W = ({\mathcal L}-\bar {\mathcal L})T$, by contradicting the characterization of dynamic discerniblity given in Proposition \ref{prop.sette}. Such proof, however, is unnecessarily involved.
\end{remark}

We now provide a first characterization of discernibility modulo synchronization. Although the derivation builds upon results available in the literature, it is developed with respect to the novel notion of discernibility introduced in this paper.
  For the sake of simplicity, we adopt the notation
\[
\mathbb{1}_k\otimes\mathbb{R}^n
:=
\{\,v\in\mathbb{R}^{kn}: v=\mathbb{1}_k\otimes z,\ \exists\,z\in\mathbb{R}^n\,\}
=
\operatorname{im}(\mathbb{1}_k\otimes I_n),
\]
and
\[
\mathbb{e}_1\otimes\mathbb{R}^n
:=
\{\,v\in\mathbb{R}^{kn}: v=\mathbb{e}_1\otimes z,\ \exists\,z\in\mathbb{R}^n\,\}
=
\operatorname{im}(\mathbb{e}_1\otimes I_n).
\]
\smallskip

\begin{proposition} \label{prop12}
  Let Assumption \ref{ass.sp.tree} hold. The following facts are equivalent.
   \begin{itemize}
       \item[i)] The systems \eqref{eq.s1} and \eqref{eq.s2} are discernible modulo synchronization.
       \item [ii)]  $X^{no}\left(\begin{bmatrix}
        I_N\otimes A-\mathcal{L}\otimes B&\\&I_N\otimes A-\mathcal{\bar L}\otimes B
    \end{bmatrix},\begin{bmatrix}
        I_{Nn}&-I_{Nn}
    \end{bmatrix}\right)
    = \mathbb{1}_{2N}\otimes \mathbb{R}^n. $
    \item[iii)] $X^{no}\left(
        I_N\otimes A-\mathcal{L}\otimes B,
       ( \mathcal{L}-\mathcal{\bar L})\otimes B\right)=\mathbb{1}_{N}\otimes \mathbb{R}^n$.
   \end{itemize} 
\end{proposition}
\begin{proof}
    $i)\Leftrightarrow ii)$ \ Follows from Definition 
    \ref{def.indiscMAS}, using the same  arguments adopted in \cite{BattistelliTesi}.

\noindent $ii)\Leftrightarrow iii)$ \ 
It is based on the fact that, for every pair of square matrices $\Phi$ and $\bar \Phi$ of the same sizes, 
    $$\begin{bmatrix} x_0 \cr x_0\end{bmatrix} \in 
    X^{no}\left(\begin{bmatrix}
        \Phi&\\&\bar \Phi
    \end{bmatrix},\begin{bmatrix}
        I_{Nn}&-I_{Nn}
    \end{bmatrix}\right) \quad \Leftrightarrow \quad x_0  \in 
    X^{no}(\Phi, \Phi-\bar \Phi).$$
\end{proof}

By making use of the Jordan form of ${\mathcal L}$ and the corresponding matrix $W=({\mathcal L}- \bar {\mathcal L})T$, we can provide alternative and more effective characterizations.

\begin{theorem} \label{theodir}
Let Assumption \ref{ass.sp.tree} holds. 
Let $0,  \lambda_2, \dots,  \lambda_r$ be the $r$ distinct eigenvalues of ${\mathcal L}$, so that the Jordan form of ${\mathcal L}$,
 $J=T^{-1} {\mathcal L} T$, is described as in \eqref{Jordan1}
 while
 $W = ({\mathcal L}-\bar {\mathcal L})T$ is block-partitioned conformably with $J$ as in \eqref{eq.W1} and \eqref{eq.W2}, with $w_1={\mathbb 0}$, i.e.,
$$W= \begin{bmatrix} \mathbb{0} & W_2 & \dots & W_r\end{bmatrix}.$$
Then the following facts are equivalent:
\begin{itemize}
\item[i)] 
The systems \eqref{eq.s1} and \eqref{eq.s2} are discernible modulo synchronization.
\item[ii)] The pair
$
\left(
\diag\left((I_{a_2}\otimes A-J_2\otimes B),\ldots, (I_{a_r}\otimes A-J_r\otimes B)\right),
\begin{bmatrix}
W_2\otimes B & \cdots & W_r\otimes B
\end{bmatrix}
\right)
$
is observable.
\item[iii)] \begin{align*}
        \rank\begin{bmatrix}
           I_{a_2}\otimes A-J_2\otimes B-\mu I&&\\&\ddots&\\
            && I_{a_r}\otimes A-J_r\otimes B-\mu I\\
            \hline
            W_2\otimes B&\cdots& W_r\otimes B
        \end{bmatrix}=(N-1)n,\quad \forall \mu\in\mathbb{C}.
    \end{align*}
    \end{itemize}
\end{theorem}

\begin{proof} $i)\Leftrightarrow ii)$\ 
    Systems \eqref{eq.s1} and \eqref{eq.s2} are discernible modulo synchronization if and only if 
   \begin{align*}   
 & X^{no}\left(
        I_N\otimes A-\mathcal{L}\otimes B,
       ( \mathcal{L}-\mathcal{\bar L})\otimes B\right)  =\mathbb{1}_{N}\otimes {\mathbb R}^n, \\ 
       \iff &  X^{no}\left((T^{-1}\otimes I_n)
        (I_N\otimes A-\mathcal{L}\otimes B)(T\otimes I_n),
       (( \mathcal{L}-\mathcal{\bar L})\otimes B)(T\otimes I_n)\right)\\
       &= X^{no}\left(
        I_N\otimes A-J\otimes B,
       \begin{bmatrix}
           \mathbb{0}& W_2 &\cdots &W_r
        \end{bmatrix}\otimes B\right)=\mathbb{e}_1\otimes \mathbb{R}^n\\
        \iff& X^{no}\left(\begin{bmatrix}A&&&\\
            & I_{a_2}\otimes A-J_2\otimes B-\mu I&&\\
            &&\ddots&\\
            &&& I_{a_r}\otimes A-J_r\otimes B-\mu I
        \end{bmatrix}, \begin{bmatrix} \mathbb{0} \otimes B &
            W_2\otimes B&\cdots &W_r\otimes B
        \end{bmatrix}\right) \\& = \mathbb{e}_{1}\otimes {\mathbb R}^n,\\
        \iff& X^{no}\left(\begin{bmatrix}
            I_{a_2}\otimes A-J_2\otimes B-\mu I&&\\
            &\ddots&\\
            && I_{a_r}\otimes A-J_r\otimes B-\mu I
        \end{bmatrix}, \begin{bmatrix}
            W_2\otimes B&\cdots &W_r\otimes B
        \end{bmatrix}\right)=\{\mathbb{0}\}.
       \end{align*}
$ii)\Leftrightarrow iii)$\ Follows immediately from the PBH observability test.
\end{proof}

 \begin{remark}  
 Theorem~\ref{theodir} extends Corollary~6 in \cite{Hao_TAC2023} in several respects. First, it applies to directed communication graphs rather than being restricted to undirected ones, while explicitly ruling out the trivial synchronization states. Moreover, it provides a simpler characterization that avoids the explicit computation of generalized eigenvectors. \end{remark}

\section{Discernibility modulo synchronization for undirected graphs}\label{secMASundir}

 In this section, we specialize the results on discernibility modulo synchronization for systems~\eqref{eq.s1} and~\eqref{eq.s2} to undirected communication graphs. 
For notational convenience, we recall here some notation that will be used in the subsequent results.
We assume that  $0 < \hat \lambda_2\le \dots \le \hat \lambda_N$ are the $N$ not necessarily distinct eigenvalues of ${\mathcal L}$, so that the Jordan form of ${\mathcal L}$ is 
 $$J=T^{-1} {\mathcal L} T = {\rm diag} (0, \hat \lambda_2, \dots, \hat \lambda_N),$$
 and  we set
 $$ W:= ({\mathcal L}-\bar {\mathcal L})T = \begin{bmatrix} {\mathbb 0} & w_2 & \dots & w_N\end{bmatrix}.$$
 We first present the counterpart of Theorem~\ref{theodir} for the undirected graph setting.

\begin{corollary} \label{cor13undir}
 Suppose that ${\mathcal G}=({\mathcal V},{\mathcal E},{\mathcal W})$ is an undirected connected graph. 
 The following facts are equivalent:
\begin{itemize}
     \item[i)] 
The systems \eqref{eq.s1} and \eqref{eq.s2} are discernible modulo synchronization.
\item[ii)] The pair
$
\left(
\diag(A-\hat\lambda_2B,\ldots,A-\hat\lambda_NB),
\begin{bmatrix}
w_2\otimes B & \cdots & w_N\otimes B
\end{bmatrix}
\right)
$
is observable.
\item[iii)] \begin{align*}
        \rank\begin{bmatrix}
            A-\hat\lambda_2B-\mu I_n&&\\&\ddots&\\
            && A-\hat \lambda_NB-\mu I_n\\
            \hline
            w_2\otimes B&\cdots& w_N\otimes B
        \end{bmatrix}=(N-1)n,\quad \forall \mu\in\mathbb{C}.
    \end{align*}
    \end{itemize}
\end{corollary}

In the case of undirected graphs, we also explore new discernibility conditions that can be derived under mild additional assumptions.

\subsection{The case of a nonsingular coupling matrix}\label{subsec.nssb}
\smallskip

We now consider the special case in which (the communication graph is undirected and) the coupling matrix $B$ is nonsingular. Although this may appear to be a restrictive assumption, it is quite natural in the present setting. Indeed, unlike in standard state-space models, the matrix $B$ does not describe the influence of external inputs on the system dynamics. Rather, it weights the disagreement between the state of each agent and those of its neighbors. Requiring $B$ to be nonsingular simply amounts to assuming that disagreements are taken into account along every state direction.

\begin{assumption}\label{ass.NSS.B}
    The coupling matrix $B \in {\mathbb R}^{n \times n}$ is nonsingular-square.
\end{assumption}

\begin{proposition} \label{prop17}
    Suppose that ${\mathcal G}=({\mathcal V},{\mathcal E},{\mathcal W})$ is an undirected connected graph. 
If Assumption \ref{ass.NSS.B} holds, then the systems \eqref{eq.s1} and \eqref{eq.s2} are discernible modulo synchronization if and only if the topology change $\mathcal{L}\rightarrow\bar{\mathcal{L}}$ is dynamically discernible.
\end{proposition}
 
\begin{proof}
    {\em Necessity:} 
Follows from Lemma \ref{LLdyndisc}.\\
  {\em Sufficiency:}    Assume, by contradiction, that the systems \eqref{eq.s1} and \eqref{eq.s2} are not discernible modulo synchronization and hence, see Corollary \ref{cor13undir},
  there exist $\mu\in\mathbb C$ and a nonzero vector
$z=
[z_2^\top\ \cdots\ z_N^\top]^\top
\in\mathbb C^{(N-1)n}$
such that
\[
\begin{bmatrix}
A-\hat\lambda_2B-\mu I_n && \\
& \ddots & \\
&& A-\hat \lambda_NB-\mu I_n\\
\hline
w_2\otimes B & \cdots & w_N\otimes B
\end{bmatrix}
z
=
{\mathbb 0} .
\]
The previous identity is equivalent to
\begin{align*}
(A-\mu I_n)z_i &= \hat \lambda_iBz_i,
\qquad &&i\in [2,N],\\
\sum_{i=2}^N(w_i\otimes B)z_i &= {\mathbb 0}.
\end{align*}
Since $B$ is nonsingular, these equations become
\begin{subequations}
\begin{align}
B^{-1}(A-\mu I_n)z_i &=  \hat\lambda_i z_i,
\qquad &&i\in [2,N],\label{cond1}\\
\sum_{i=2}^N(w_i\otimes Bz_i)&={\mathbb 0}.
\label{cond2}
\end{align}
\end{subequations}
Condition \eqref{cond2} can be equivalently written as
\[
\sum_{i=2}^N[w_i]_k\,Bz_i={\mathbb 0},
\qquad
\forall\ k\in[1,N],
\]
and hence as 
\[
\begin{bmatrix}
Bz_{2}&\cdots&Bz_{N}
\end{bmatrix}
\begin{bmatrix}
w_{2}^\top\\
\vdots\\
w_{N}^\top
\end{bmatrix}
=
{\mathbb 0}.
\]
Using again the nonsingularity of $B$, we obtain
\[
\begin{bmatrix}
 z_{2}&\cdots&z_{N}
\end{bmatrix}
\begin{bmatrix}
w_{2}^\top\\
\vdots\\
w_{N}^\top
\end{bmatrix}
=
{\mathbb 0}.
\]
By \eqref{cond1}, each vector $z_{i}$ is an eigenvector of
$B^{-1}(A-\mu I_n)$ associated with the eigenvalue $ \hat\lambda_{i}$.
By grouping together the  eigenvectors corresponding to the same distinct eigenvalue, the above identity can be written as
\[
\begin{bmatrix}
Z_2&\cdots&Z_r
\end{bmatrix}
\begin{bmatrix}
 W_2^\top\\
\vdots\\
 W_r^\top
\end{bmatrix}
=
{\mathbb 0},
\]
where the columns of each block $Z_\ell$ are  eigenvectors associated with the same eigenvalue  $\lambda_\ell \in \{\lambda_2, \dots, \lambda_r\}= \{\hat \lambda_2, \dots, \hat 
\lambda_N\}$ of $B^{-1}(A-\mu I_n)$. 

Since not every block $Z_\ell$ is zero, Lemma~\ref{techlemma} implies that at least one of the matrices $W_\ell, \ell\in [2,r]$, has not full column rank. By Corollary~\ref{corotto}, it follows that the topology change
${\mathcal L}\rightarrow\bar{\mathcal L}$ is not dynamically discernible. 
\end{proof}
 \smallskip

\subsection{Conditions on the spectra of the matrices $A-\hat\lambda_iB$}\label{subsec.spectralcond}  
\smallskip

We now investigate discernibility under suitable assumptions on the spectra of the matrices $A-\hat{\lambda}_iB$, $i\in[2,N]$. A first characterization of the discernibility of systems~\eqref{eq.s1} and~\eqref{eq.s2} based on such spectral assumptions was proposed in \cite{Roy_TCNS2020} to correct Proposition~2 in \cite{BattistelliTesi}. Specifically, \cite{Roy_TCNS2020} assumes that, for every pair of distinct Laplacian eigenvalues $\hat{\lambda}_i$ and $\hat{\lambda}_j$, the matrices $A-\hat{\lambda}_iB$ and $A-\hat{\lambda}_jB$ have disjoint spectra.

Theorem~\ref{prop.disc.mas.genB} and Corollary~\ref{cor.disc.mas.genB} show that discernibility modulo synchronization is equivalent to the dynamical discernibility of the corresponding topology change under significantly weaker spectral assumptions.

 \begin{theorem}\label{prop.disc.mas.genB}
    Suppose that ${\mathcal G}=({\mathcal V},{\mathcal E},{\mathcal W})$ is an undirected connected graph.
Assume that for every $\mu\in\bigcap_{i=2}^N \sigma(A-\hat \lambda_i B),$ there exists $\epsilon\in\mathbb{R}$ such that $\mu\notin\sigma(A-\epsilon B)$.
Then the systems \eqref{eq.s1} and \eqref{eq.s2} are discernible modulo synchronization if and only if the topology change $\mathcal{L}\rightarrow\bar{\mathcal{L}}$ is dynamically discernible. 
\end{theorem}
\begin{proof}
{\em Necessity:} Follows from Lemma \ref{LLdyndisc}.\smallskip

 {\em Sufficiency:} Suppose that for every $\mu\in\bigcap_{i=2}^N \sigma(A-\hat \lambda_i B),$ there exists $\epsilon\in\mathbb{R}$ such that $\mu\notin\sigma(A-\epsilon B)$.  By Lemma \ref{techlemma2} in the Appendix, this implies that there exists $\bar \epsilon\in {\mathbb R}$ such that $\sigma(A-\bar\epsilon B)\cap \left(\bigcup_{i=2}^N\sigma(A-\hat \lambda_i B)\right)=\emptyset$.\\
  Assume, now, that there exist $\mu\in\mathbb C$ and a nonzero vector
\[
z=
[z_2^\top\ \cdots\ z_N^\top]^\top
\in\mathbb C^{(N-1)n}
\]
such that
\[
\begin{bmatrix}
A- \hat\lambda_2 B-\mu I_n && \\
& \ddots & \\
&& A- \hat\lambda_NB-\mu I_n\\
\hline
w_2\otimes B & \cdots & w_N\otimes B
\end{bmatrix}
z
=
{\mathbb 0}.
\]
Note that $\mu$ necessarily belongs to $\bigcup_{i=2}^N\sigma(A-\hat \lambda_i B)$.
The  previous identity is equivalent to
\begin{subequations}\label{eq.cond.B.sing}
\begin{align}
(A-\mu I_n)z_i &=  \hat\lambda_i Bz_i,
\qquad &&i\in [2,N],\label{eq.cond.B.sing.1}\\
\sum_{i=2}^N(w_i\otimes B)z_i &= {\mathbb 0}.\label{eq.cond.B.sing.2}
\end{align}
\end{subequations}
By subtracting  $\bar \epsilon Bz_i$ on both sides of  \eqref{eq.cond.B.sing.1}, we obtain
\begin{align}
    (A-\bar\epsilon B-\mu I_n)z_i &= ( \hat\lambda_i-\bar\epsilon) Bz_i,
\qquad &&i\in [2,N].
\label{eq.cicci}
\end{align}
We note that, since $\sigma(A-\bar\epsilon B)\cap \left(\bigcup_{i=2}^N\sigma(A-\hat \lambda_i B)\right)=\emptyset$, condition $\hat\lambda_i-\bar \epsilon \ne 0$ holds for every $i\in [2,N]$. Also, since  $\mu\notin\sigma(A-\bar\epsilon B)$, the matrix $A-\bar\epsilon B-\mu I_n$ is nonsingular, and hence \eqref{eq.cicci} can be rewritten as
\begin{align}
    (A-\bar\epsilon B-\mu I_n)^{-1} B z_i &= \frac{1}{ \hat\lambda_i-\bar\epsilon} z_i,
\qquad &&i\in [2,N],
\label{eq.cicci2}
\end{align}
which shows that all vectors $z_i$ are eigenvectors of the same matrix $(A-\bar\epsilon B-\mu I_n)^{-1} B$ corresponding to (not necessarily distinct) eigenvalues $\frac{1}{\hat\lambda_i-\bar\epsilon}$.
By making use of \eqref{eq.cicci}, we can also write
\begin{align}\label{eq.def.Bzi}
Bz_i=\frac{1}{( \hat\lambda_i-\bar\epsilon) }(A-\bar\epsilon B-\mu I_n)z_i,
\qquad &&i\in [2,N].
\end{align}
On the other hand, condition \eqref{eq.cond.B.sing.2} is equivalent to
\[
\sum_{i=2}^N[w_i]_k\,Bz_i={\mathbb 0},
\qquad
\forall\ k\in[1,N],
\]
or, equivalently,
\begin{align}\label{eq.BzW=0}
    \begin{bmatrix}
Bz_{2}&\cdots&Bz_{N}
\end{bmatrix}
\begin{bmatrix}
w_{2}^\top\\
\vdots\\
w_{N}^\top
\end{bmatrix}
=
{\mathbb 0}.
\end{align}
Substituting \eqref{eq.def.Bzi} in \eqref{eq.BzW=0}, and collecting the matrix $(A-\bar\epsilon B-\mu I_n)$ on the left, we obtain
\begin{align}\label{eq.zw.B.sing}
(A-\bar\epsilon B-\mu I_n)\begin{bmatrix}
\frac{1}{\hat\lambda_{2}-\bar\epsilon}z_{2}&\cdots&\frac{1}{\hat\lambda_{N}-\bar\epsilon}z_{N}
\end{bmatrix}
\begin{bmatrix}
w_{2}^\top\\
\vdots\\
w_{N}^\top
\end{bmatrix}
=
{\mathbb 0}.
\end{align}
Since  the matrix $(A-\bar\epsilon B-\mu I_n)$ is nonsingular square, equation \eqref{eq.zw.B.sing} holds if and only if
\[
\begin{bmatrix}
\frac{1}{\hat\lambda_{2}-\bar\epsilon}z_{2}&\cdots&\frac{1}{\hat\lambda_{N}-\bar\epsilon}z_{N}
\end{bmatrix}
\begin{bmatrix}
w_{2}^\top\\
\vdots\\
w_{N}^\top
\end{bmatrix}
=
{\mathbb 0}.
\]
Grouping together the eigenvectors $\frac{1}{\hat\lambda_{i}-\bar\epsilon}z_{i}, i\in [2,N],$ of $(A-\bar\epsilon B-\mu I_n)^{-1}B$ corresponding to the same distinct eigenvalue $\lambda_\ell\in \{\lambda_2, \dots, \lambda_r\}=\{\hat \lambda_2, \dots, \hat \lambda_N\}$, the above identity can be written as
\[
\begin{bmatrix}
Z_2&\cdots&Z_r
\end{bmatrix}
\begin{bmatrix}
W_2^\top\\
\vdots\\
W_r^\top
\end{bmatrix}
=
{\mathbb 0}.
\]
Since not every block $Z_i$ is zero, by resorting to the same reasoning as in the proof of Proposition \ref{prop17}, we can conclude that the dynamic discernibility of the topology change ${\mathcal L} \to \bar {\mathcal L}$ is contradicted.
\end{proof}

\begin{remark} Note that the additional hypothesis introduced in Theorem \ref{prop.disc.mas.genB},
that for every $\mu\in\bigcap_{i=2}^N \sigma(A-\hat \lambda_i B),$ there exists $\epsilon\in\mathbb{R}$ such that $\mu\notin\sigma(A-\epsilon B)$, necessarily implies that $(A,B)$ is observable, as shown in Lemma
    \ref{lem.AB.obs.necessary}, in the Appendix.
\end{remark}

 \begin{corollary}\label{cor.disc.mas.genB}
    Suppose that ${\mathcal G}=({\mathcal V},{\mathcal E},{\mathcal W})$ is an undirected connected graph.
If  $\bigcap_{i=2}^N \sigma(A-\hat \lambda_i B)=\emptyset$, 
then the systems \eqref{eq.s1} and \eqref{eq.s2} are discernible modulo synchronization if and only if the topology change $\mathcal{L}\rightarrow\bar{\mathcal{L}}$ is dynamically discernible. 
\end{corollary}
\begin{proof}
The proof follows the same arguments as the previous one, the only difference being that it relies on Corollary~\ref{cortechlemma2} in the Appendix instead of Lemma~\ref{techlemma2}. 
\end{proof}
\smallskip

 \subsection{Single edge disconnection or change of weight}\label{subsec.MASedgedisc}
\smallskip

We now investigate the discernibility problem for MASs with an undirected, connected communication graph under either the disconnection   or a variation in the weight of an existing edge.  Assume that the topology variation occurs between nodes $i$ and $j$. As established in Lemma \ref{LLdyndisc}, a necessary condition for the two systems \eqref{eq.s1} and \eqref{eq.s2} to be discernible modulo synchronization is that the topology change $\mathcal{L}\rightarrow\bar{\mathcal{L}}$ be dynamically discernible. By Proposition \ref{ED_eigenvectorentries} and Remark \ref{remark10}, this implies that $\mathcal{L}$ has $N$ distinct and real eigenvalues $\{0,\lambda_2, \dots, \lambda_N\}$ and that, for every (nonzero) eigenvalue $\lambda_\ell\in\sigma(\mathcal{L})$, $\ell\in[2,N]$, the corresponding eigenvector $v_\ell$ satisfies $[v_\ell]_i\neq [v_\ell]_j$.
In the following proposition, we further strengthen the result of Theorem \ref{prop.disc.mas.genB} for the particular case of a single-edge disconnection. 
  The result relies on the assumption that \(N\ge n+2\), which is generally reasonable, as the number of agents typically far exceeds the dimension of the individual agent model.

\begin{proposition}\label{prop.edge.disc.mas}
  Suppose that ${\mathcal G}=({\mathcal V},{\mathcal E},{\mathcal W})$ is an undirected connected graph, and assume that $n+2\le N$. Let the topology change
$\mathcal{L}\rightarrow\bar{\mathcal{L}}$ resulting from a variation in the weight of the edge
$(j,i)\in{\mathcal E}$ be dynamically discernible  (which implies that the eigenvalues of ${\mathcal L}$ are all distinct).  
   Then the two dynamical systems \eqref{eq.s1} and \eqref{eq.s2} are discernible modulo synchronization if and only if for every $\mu\in\mathbb{C}$
   \begin{equation}
   \left| \left\{i\in[2,N]\ :\ \mu\in\sigma(A-\lambda_iB)\right\}\right|\le n.
       \label{eq.cardn}
   \end{equation}
\end{proposition}
\begin{proof}
{\em Necessity:} Assume, by contradiction, that there exists $\mu\in\mathbb C$ such that 
$$\left| \left\{i\in[2,N]\ :\ \mu\in\sigma(A-\lambda_iB)\right\}\right|>n.$$
Consequently,
$$\mathcal{I} := \left\{i\in[2,N]\ :\ \mu\in\sigma(A-\lambda_iB)\right\} = \{ i_1, i_2, \dots, i_k\},$$
with $k > n$. Let
$z_{i_h}$ be a nonzero eigenvector of $(A - \lambda_{i_h} B)$ corresponding to the eigenvalue $\mu$. Clearly, for every $c_h \ne 0$, 
\begin{align*}
    (A-\lambda_{i_h}B -\mu I_n)c_hz_{i_h}=\mathbb{0}.
\end{align*}
We want to prove that it is always possible to choose the coefficients $c_h, h\in [1,k],$
in such a way that 
$$\sum_{h=1}^k (w_{i_h} \otimes B)c_h z_{i_h} = {\mathbb 0}.$$
The previous condition can be equivalently rewritten as
\begin{equation}
{\mathbb 0} = \sum_{h=1}^k (w_{i_h} \otimes c_h B z_{i_h}) = \begin{bmatrix}
        Bz_{i_1}& Bz_{i_2}&\cdots& Bz_{i_k}
    \end{bmatrix}\begin{bmatrix}
        c_{i_1}&&\\
        &\ddots&\\
        &&c_{i_k}\
    \end{bmatrix}\begin{bmatrix}
        w_{i_1}^\top\\
        w_{i_2}^\top\\
        \vdots\\
        w_{i_k}^\top
    \end{bmatrix}.
\label{cond.ci}
\end{equation}
Now we recall that the disconnection (or weight variation) of the edge $(j,i)$ is modeled as
\[
{\mathcal L}-\bar{\mathcal L}
=
\alpha({\mathbb e}_i-{\mathbb e}_j)
({\mathbb e}_i-{\mathbb e}_j)^\top,
\]
for some $\alpha \ne 0$. Consequently, for every $h\in [1,k]$,
$$w_{i_h} = ({\mathcal L}-\bar{\mathcal L}) T {\mathbb e}_{i_h} = \alpha({\mathbb e}_i-{\mathbb e}_j)
({\mathbb e}_i-{\mathbb e}_j)^\top T {\mathbb e}_{i_h} 
= \alpha({\mathbb e}_i-{\mathbb e}_j) ([v_{i_h}]_i- [v_{i_h}]_j).$$
This implies that equation \eqref{cond.ci}, can be written as
$$
{\mathbb 0}= \begin{bmatrix}
        Bz_{i_1}& Bz_{i_2}&\cdots& Bz_{i_k}
    \end{bmatrix}\begin{bmatrix}
        c_{i_1}&&\\
        &\ddots&\\
        &&c_{i_k}
    \end{bmatrix}\begin{bmatrix}
        [v_{i_1}]_i-[v_{i_1}]_j\\
        [v_{i_2}]_i-[v_{i_2}]_j\\
        \vdots\\
        [v_{i_k}]_i-[v_{i_k}]_j
    \end{bmatrix}\alpha (\mathbb{e}_i - {\mathbb e}_j)^\top
$$
which holds true if and only if
$$
{\mathbb 0}= \begin{bmatrix}
        Bz_{i_1}& Bz_{i_2}&\cdots& Bz_{i_k}
    \end{bmatrix}\begin{bmatrix}
        c_{i_1}&&\\
        &\ddots&\\
        &&c_{i_k}
    \end{bmatrix}\begin{bmatrix}
        [v_{i_1}]_i-[v_{i_1}]_j\\
        [v_{i_2}]_i-[v_{i_2}]_j\\
        \vdots\\
        [v_{i_k}]_i-[v_{i_k}]_j
    \end{bmatrix}.
$$
Since the matrix $\begin{bmatrix}
        Bz_{i_1}& Bz_{i_2}&\cdots& Bz_{i_k}
    \end{bmatrix}$ has size $n \times k$ and $k > n$, its kernel is nontrivial. This means that 
    there exists $[\alpha_1 \ \alpha_2 \ \dots \ \alpha_k]^\top \ne {\mathbb 0}$ belonging to its kernel.
    Clearly, by the  hypothesis of dynamical discernibility of 
${\mathcal L}\to \bar {\mathcal L}$, 
 every $( [v_{i_h}]_i-[v_{i_h}]_j)$ is nonzero, consequently the equation
 $$\begin{bmatrix}
         c_{i_1} ([v_{i_1}]_i-[v_{i_1}]_j)\\
       c_{i_2}  ( [v_{i_2}]_i-[v_{i_2}]_j)\\
        \vdots\\
       c_{i_k}  ( [v_{i_k}]_i-[v_{i_k}]_j)
    \end{bmatrix} =
    \begin{bmatrix} \alpha_1 \cr \alpha_2 \cr \vdots \cr \alpha_k\end{bmatrix},$$
    in the unknown coefficients $c_h, h\in [1,k],$ is solvable, and hence there exist vectors 
    $c_{h} z_{i_h}, h\in [1,k],$ not all zero, such that
 \begin{subequations}\label{eq.cond.edge.mas}
\begin{align}
(A-\mu I_n)c_hz_{i_h} &= \lambda_{i_h}B c_hz_{i_h},
\qquad &&h\in [1,k],\label{eq.cond.edge.mas.1}\\
\sum_{h=1}^k(w_{i_h}\otimes B)c_hz_{i_h} &= {\mathbb 0}.\label{eq.cond.edge.mas.2}
\end{align}
\end{subequations}
This ensures that there exists a nonzero vector in the kernel of 
\[
\begin{bmatrix}
A-\lambda_{2} B-\mu I_n && \\
& \ddots & \\
&& A-\lambda_NB-\mu I_n\\
\hline
w_2\otimes B & \cdots & w_N\otimes B
\end{bmatrix}
\]
and this implies that 
 the two dynamical systems \eqref{eq.s1} and \eqref{eq.s2} are not discernible modulo synchronization.
 \smallskip
 
{\em Sufficiency:}
We preliminarily observe that, by hypothesis, $n+2\le N$, and therefore $n<N-1$. Hence, for every $\mu\in\mathbb{C}$, condition \eqref{eq.cardn} becomes
\begin{align*}
\left|\left\{i\in[2,N]:\mu\in\sigma(A-\lambda_iB)\right\}\right|
\le n<N-1.
\end{align*}
Consequently, 
\[\bigcap_{i=2}^N\sigma(A- \lambda_i B) =\emptyset.\]
This ensures, by Corollary \ref{cor.disc.mas.genB}, that  the two dynamical systems \eqref{eq.s1} and \eqref{eq.s2} are discernible modulo synchronization.
\end{proof}

\begin{example}
Consider the discrete-time system
\begin{align*}
     x(t+1)=(I_N\otimes A-\mathcal{L}\otimes B)x(t)
\end{align*}
where
\begin{align*}
A &=
\begin{bmatrix}
1 & 1\\
0 & 0.5
\end{bmatrix},
\quad
B =
\begin{bmatrix}
1 & 0\\
0 & 0
\end{bmatrix},\quad  \mathcal{L} =
\begin{bmatrix}
3 & -1 & 0 & -1 & -1 \\
-1 & 2 & -1 & 0 & 0 \\
0 & -1 & 2 & 0 & -1 \\
-1 & 0 & 0 & 2 & -1 \\
-1 & 0 & -1 & -1 & 3
\end{bmatrix}
\end{align*}
and $\mathcal{L}$ is the Laplacian matrix associated with the undirected graph $\mathcal{G}$ represented in Figure \ref{fig:topology-change} (on the  left). Suppose that a topology change occurs as a consequence of the disconnection of edge $(1,2)$. The dynamics of the resulting system  (whose associated graph is on the right of Figure \ref{fig:topology-change}) is described by
\begin{align*}
  \bar   x(t+1)=(I_N\otimes A-\mathcal{\bar L}\otimes B)\bar x(t),\quad
\text{where} \quad \bar{\mathcal{L}}
=
\begin{bmatrix}
2 & 0 & 0 & -1 & -1 \\
0 & 1 & -1 & 0 & 0 \\
0 & -1 & 2 & 0 & -1 \\
-1 & 0 & 0 & 2 & -1 \\
-1 & 0 & -1 & -1 & 3
\end{bmatrix}.
\end{align*}

\begin{figure}
\begin{center}
\begin{tikzpicture}[
    node/.style={
        circle,
        draw,
        minimum size=9mm,
        inner sep=0pt
    },
    edge/.style={
        thick
    },
    >=stealth
]


\node[node] (1) at (0,2) {$1$};
\node[node] (2) at (1.8,3.2) {$2$};
\node[node] (3) at (3.6,2) {$3$};
\node[node] (5) at (2.9,0) {$5$};
\node[node] (4) at (0.7,0) {$4$};

\draw[edge] (1) -- (2);
\draw[edge] (1) -- (4);
\draw[edge] (1) -- (5);
\draw[edge] (2) -- (3);
\draw[edge] (3) -- (5);
\draw[edge] (4) -- (5);

\node at (1.8,4) {$\mathcal{G}$};


\draw[->, very thick]
    (4.5,1.6) -- (6.8,1.6)
    node[midway, above, align=center]
    {edge $(1,2)$\\ disconnection};


\node[node] (b1) at (7.7,2) {$1$};
\node[node] (b2) at (9.5,3.2) {$2$};
\node[node] (b3) at (11.3,2) {$3$};
\node[node] (b5) at (10.6,0) {$5$};
\node[node] (b4) at (8.4,0) {$4$};

\draw[edge] (b1) -- (b4);
\draw[edge] (b1) -- (b5);
\draw[edge] (b2) -- (b3);
\draw[edge] (b3) -- (b5);
\draw[edge] (b4) -- (b5);

\node at (9.5,4) {$\bar{\mathcal{G}}$};

\end{tikzpicture}
\end{center}
\caption{Topology change due to the disconnection of edge $(1,2)$.}
\label{fig:topology-change}
\end{figure}
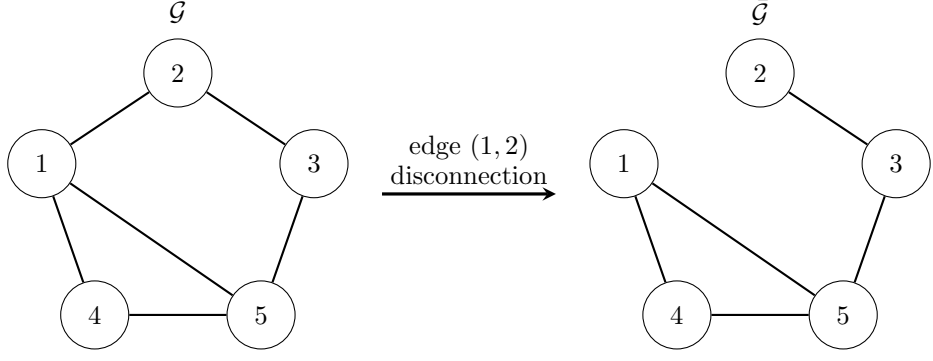
We aim to determine whether the topology change is discernible modulo synchronization. As a preliminary step, we verify that the two necessary conditions derived in Lemmas \ref{ABobs} and \ref{LLdyndisc} hold. We easily see that the pair $(A,B)$ is observable.
Since the considered topology modification consists of a single-edge disconnection,
to check 
 that the topology change $\mathcal{L}\rightarrow\mathcal{\bar L}$ is dynamically discernible we can apply Proposition \ref{ED_eigenvectorentries} and Remark \ref{remark10}.
 This means that we have to verify that ${\mathcal L}$ has $5$ distinct eigenvalues and that for each of the 4 eigenvectors associated with the nonzero eigenvalues the first two entries do not coincide.
 The spectrum of ${\mathcal L}$ is
 $$\sigma({\mathcal L})= \{\lambda_1, \lambda_2, \lambda_3, \lambda_4, \lambda_5\}=\{0,1.3820, 2.3820, 3.6180, 4.6180\}$$
 and hence the first constraint is satisfied.
 A nonsingular transformation matrix that reduces ${\mathcal L}$ to Jordan form 
\[J = T^{-1} \mathcal{L}T
=
\begin{bmatrix}
0 & 0 & 0 & 0 & 0 \\
0 & 1.3820 & 0 & 0 & 0 \\
0 & 0 & 2.3820 & 0 & 0 \\
0 & 0 & 0 & 3.6180 & 0 \\
0 & 0 & 0 & 0 & 4.6180
\end{bmatrix},\]
is 
\begin{align*}
T =
\begin{bmatrix}
-0.4472 & -0.1954 & 0.3717 & -0.5117 & 0.6015 \\
-0.4472 & 0.5117 & 0.6015 & 0.1954 & -0.3717 \\
-0.4472 & 0.5117 & -0.6015 & 0.1954 & 0.3717 \\
-0.4472 & -0.6325 & 0 & 0.6325 & 0 \\
-0.4472 & -0.1954 & -0.3717 & -0.5117 & -0.6015
\end{bmatrix},
\end{align*}
whose columns are normalized eigenvectors of $\mathcal{L}$. 
It is immediate to verify that  also condition $ii)$ of Proposition \ref{ED_eigenvectorentries} is satisfied, since the last $4$ columns have the first and the second entries that do not coincide. Therefore, the topology change $\mathcal{L}\rightarrow\mathcal{\bar L}$ is dynamically discernible. 
We can now proceed with the analysis of the discernibility modulo synchronization.
Instead of resorting to Corollary \ref{cor13undir}, that is computationally more demanding, 
we try to see if any of the other results derived for the case of undirected graphs can be applied. We first observe that since $B$ is singular we cannot make use of Proposition \ref{prop17}. 
We now   evaluate whether the hypotheses of Theorem \ref{prop.disc.mas.genB} hold. 
To this end, we evaluate the spectra of the matrices
$A-\lambda_i B$, for $i\in [2,5]$.
Since the matrix
\begin{align*}
A-\lambda_i B
&=
\begin{bmatrix}
1-\lambda_i & 1\\
0 & 0.5
\end{bmatrix}
\end{align*}
is upper triangular, its spectrum is immediately identified as
$\sigma(A-\lambda_i B)
=
\left\{
1-\lambda_i,\,
3
\right\}.$
Consequently, 
\begin{align*}
\sigma(A-\lambda_2B)=
\{-0.3820,0.5\},
\quad
\sigma(A-\lambda_3B)=
\{-1,3820,0.5\},\\
\sigma(A-\lambda_4B)=
\{-2.6180,0.5\},\quad
\sigma(A-\lambda_5B)=
\{-3.6180,0.5\}.
\end{align*}
Thus, the eigenvalue $0.5$ is common to all the matrices $A-\lambda_iB$, namely,
$
0.5
\in
\bigcap_{i=2}^{5}
\sigma(A-\lambda_iB),
$
and no $\epsilon\in {\mathbb R}$ can be found such that $0.5\not\in  \sigma(A - \epsilon B)$.
Therefore, Theorem \ref{prop.disc.mas.genB} cannot be used to establish discernibility modulo synchronization. Nevertheless, since $n+2=N$ and the topology change $\mathcal{L}\rightarrow\mathcal{\bar L}$, associated with the edge disconnection, is dynamically discernible, we can apply Proposition
\ref{prop.edge.disc.mas} stating  that, under these hypotheses,  the two multi-agent systems are discernible modulo synchronization if and only if, for every $\mu\in\mathbb{C}$,
   $$\left| \left\{i\in[2,5]\ :\ \mu\in\sigma(A-\lambda_iB)\right\}\right|\le 2.$$
For $\mu=0.5$, however, one sees that
 $$\left| \left\{i\in[2,5]\ :\ 0.5\in\sigma(A-\lambda_iB)\right\}\right|=4 >  2$$
and therefore the condition of Proposition \ref{prop.edge.disc.mas} is violated.
We conclude that, although the topology change $\mathcal{L}\rightarrow\mathcal{\bar L}$ is dynamically discernible, the corresponding multi-agent systems are not discernible modulo synchronization.

For example, consider the case in which each agent of the original and modified model is initialized with the following initial conditions
\begin{align*}
x_1(0) =\bar x_1(0) =
\begin{bmatrix}
 0.5176\\
   -3.2890
\end{bmatrix},
\quad
x_2(0) =\bar x_2(0) =&
\begin{bmatrix}
 0.5176\\
    1.3231
\end{bmatrix},
\quad
x_3(0) =\bar x_3(0) =
\begin{bmatrix}
  -0.8403\\
   -2.3484
\end{bmatrix},
\\
x_4(0) =\bar x_4(0) =
\begin{bmatrix}
 3.4951\\
    4.1548
\end{bmatrix}&,
\quad
x_5(0) =\bar x_5(0) =
\begin{bmatrix}
 0.7942\\
   -0.9629
\end{bmatrix},
\end{align*}
then from Figure~\ref{fig:indiscernibility-trajectories} it is immediate to verify that the two systems generate the same state trajectories.
\begin{figure}
    \centering
    \includegraphics[width=0.32\linewidth]{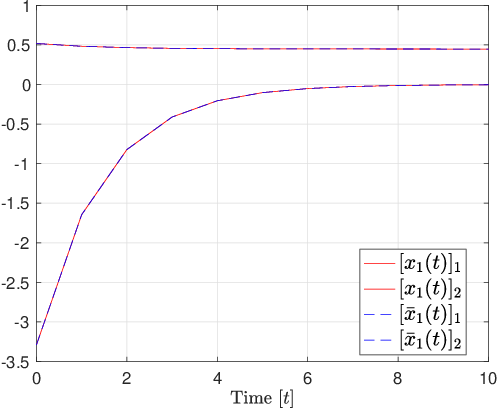}
    \includegraphics[width=0.32\linewidth]{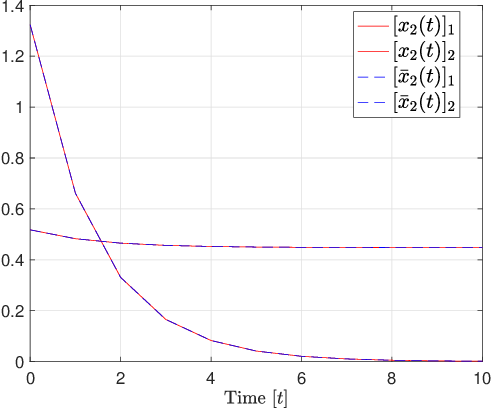}
    \includegraphics[width=0.32\linewidth]{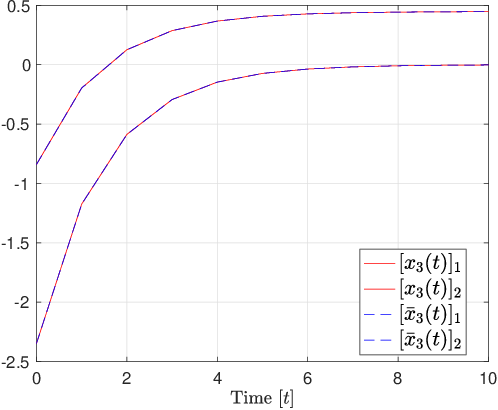}
   \\[1em]
    \includegraphics[width=0.32\linewidth]{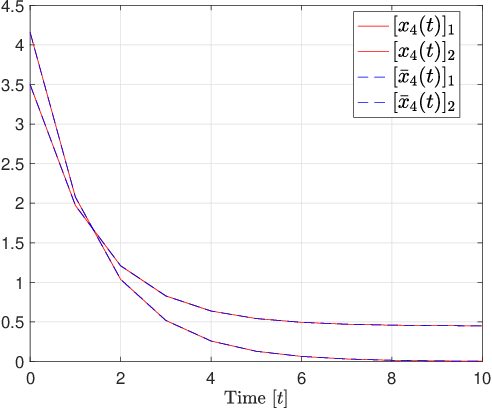}
    \includegraphics[width=0.32\linewidth]{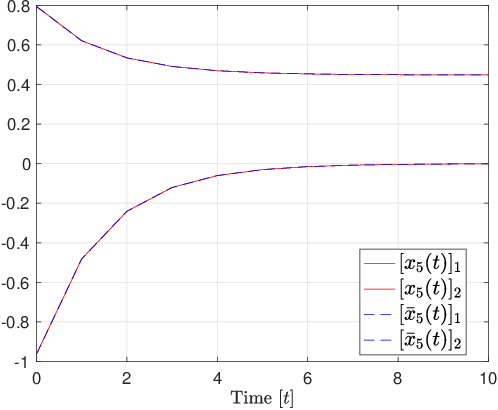}
    \caption{State trajectories of the five agents for the original system and for the modified system obtained by disconnecting edge $(1,2)$. Each subplot shows the evolution of the two state components of the corresponding agent.}
    \label{fig:indiscernibility-trajectories}
\end{figure}
\end{example}

\section{Conclusions}\label{sec.conclusions}
 
In this article, we introduced and characterized a notion of discernibility specifically tailored to the detection of topology changes in networked dynamical systems.

We first considered elementary consensus networks, whose dynamics is governed by the Laplacian matrix ${\mathcal L}$ in continuous time and by the matrix $I-\alpha{\mathcal L}$ in discrete time. To eliminate the intrinsic indiscernibility associated with the consensus subspace, we introduced the notion of \emph{dynamical discernibility of a topology change}. We derived a complete characterization for directed communication graphs and subsequently specialized the analysis to edge disconnections, obtaining conditions that admit a direct interpretation in terms of the eigenvectors of the original Laplacian matrix. We also derived simplified characterizations for undirected communication graphs.

In the second part of the paper, we extended the analysis to homogeneous multi-agent systems with higher-order dynamics described by the pair of matrices $(A,B)$. Since synchronization trajectories constitute an unavoidable source of indiscernibility in such systems, we introduced the notion of \emph{discernibility modulo synchronization}, which explicitly excludes this trivial ambiguity. We first established necessary conditions and then derived necessary and sufficient conditions for discernibility modulo synchronization. For undirected communication graphs, we obtained several simplified characterizations under mild additional assumptions. In particular, we showed that, when the coupling matrix $B$ is nonsingular, dynamical discernibility of the topology change is equivalent to discernibility modulo synchronization. We also derived further characterizations under suitable assumptions on the spectra of the matrices $A-\lambda B$, where $\lambda$ denotes a generic eigenvalue of the Laplacian ${\mathcal L}$. Finally, specializing the analysis to edge disconnections, we obtained necessary and sufficient conditions for discernibility modulo synchronization under the assumption of dynamical discernibility of the topology change.

All the results established for undirected communication graphs extend immediately to directed graphs with diagonalizable Laplacian matrices. Extending the analysis to general directed graphs with non-diagonalizable Laplacians is considerably more challenging and constitutes an interesting direction for future research.

Other natural research directions include extending the proposed framework to broader classes of topology variations and deriving simpler necessary and sufficient conditions for discernibility. Another promising direction is the extension of the present framework to output discernibility and $M$-discernibility, where only partial state measurements, or measurements of selected components of the agent state, are available.

\appendix
\section{Appendix}

 This appendix collects the technical lemmas used throughout the paper whose proofs or statements, while instrumental to the development of the main results, are not central to the paper's primary contribution.

     \begin{lemma} \label{techlemma}
 Let $M \in {\mathbb R}^{n \times n}$ be a square matrix, let $\lambda_1,  \lambda_2, \dots,   \lambda_\delta\in {\mathbb C}$ be $\delta$ distinct eigenvalues of $M$.
 Let $Z_i\in {\mathbb C}^{n\times n_i}, i\in [1,\delta],$ be a matrix whose columns are eigenvectors of $M$ corresponding to $\lambda_i$, and let $W_i\in {\mathbb R}^{p \times n_i}, i\in [1,\delta],$ be a full column rank matrix. Then, condition
 \begin{equation}
 \begin{bmatrix} Z_1 & Z_2 & \dots & Z_\delta\end{bmatrix}  \begin{bmatrix} W_1^\top \cr W_2^\top \cr \vdots \cr W_\delta^\top\end{bmatrix}  = {\mathbb 0}_{n\times p}
 \label{VW0}
 \end{equation}
 implies $Z_i=0$, for every $i\in [1,\delta]$.
 \end{lemma}
 
 \begin{proof}
 Condition \eqref{VW0} implies that for every $j\in [1,p]$
 $$\begin{bmatrix} Z_1 & Z_2 & \dots & Z_\delta\end{bmatrix}  \begin{bmatrix} W_1^\top \cr W_2^\top \cr \vdots \cr W_\delta^\top\end{bmatrix}  {\mathbb e}_j = {\mathbb 0}_n.$$
So, if we set $q_i^{(j)} := 
W_i^\top {\mathbb e}_j$, we get
 $$\sum_{i=1}^\delta Z_i q_i^{(j)} = {\mathbb 0}_n.$$
 Since $Z_i q_i^{(j)} $ is an eigenvector of $M$ corresponding to the eigenvalue $\lambda_i$ and the eigenvectors of $M$ corresponding to distinct eigenvalues are linearly independent, it follows that
 $Z_i q_i^{(j)} = {\mathbb 0}_n$ for every $i\in[1,\delta].$ On the other hand, since this results is true for every $j\in [1,p]$, it follows that
 $$Z_i W_i^\top = {\mathbb 0}_{n \times p},$$
 and since $W_i^\top$ is of full row rank, it follows that $Z_i={\mathbb 0}_{n\times n_i}$. 
 This implies that $Z_i= {\mathbb 0}$ for every $i\in [1,\delta]$.
\end{proof}

 \begin{lemma}\label{techlemma2}
Given two $n\times n$ real matrices $A$ and $B$ and a family of (not necessarily distinct) complex numbers $\{\hat\lambda_2, \hat\lambda_3, \dots, \hat \lambda_N\}$,
set
$$\{\mu_1, \mu_2, \dots, \mu_d\} := \bigcap_{i=2}^N\sigma(A-\hat \lambda_i B),$$
and 
$$\{\mu_1, \mu_2, \dots, \mu_d, \mu_{d+1}, \dots, \mu_D\} := \bigcup_{i=2}^N\sigma(A-\hat \lambda_i B).$$
    The following facts are equivalent.
    \begin{itemize}
        \item[i)] For every $\mu_j, j\in [1,d]$, there exists $\epsilon_j \in {\mathbb R}$
        such that $\mu_j\notin \sigma(A-\epsilon_j B)$.
        \item[ii)] There exists a real number $\bar \epsilon \in {\mathbb R}$ such that $\sigma(A-\bar\epsilon B)\cap  \{\mu_1, \mu_2, \dots, \mu_d, \mu_{d+1}, \dots, \mu_D\} =\emptyset$.
    \end{itemize}
    \end{lemma}
   \begin{proof}
$i) \Rightarrow ii)$ \ 
Suppose that, for each $j\in[1,d]$, there exists $\epsilon_j\in {\mathbb R},$ such that
$\mu_j\notin\sigma(A-\epsilon_j B)$.  
Then
$
\det(A-\mu_j I-\epsilon_j B)\neq 0,
$
and therefore the polynomial
$
p_j(x):=\det(A-\mu_j I-xB)
$
is not identically zero.
Set
$
{\mathcal Z}_j:=\{x\in\mathbb C:p_j(x)=0\}.
$
Since $p_j(x)$ is a nonzero polynomial, ${\mathcal Z}_j$ is a finite set. Moreover,
\begin{align*}
\mu_j\notin\sigma(A-xB)
\quad\Longleftrightarrow\quad
x\notin {\mathcal Z}_j.
\end{align*}
Consider, now, the other eigenvalues $\mu_j, j\in [d+1,D].$ For each of them we know that there exists 
$\hat \lambda_i, i\in [2,N]$, such that $\mu_j \not\in
\sigma(A-\hat \lambda_i B)$. This ensures, also in this case, that the polynomial
$
p_j(x):=\det(A-\mu_j I-xB)
$
is not identically zero and that
$
{\mathcal Z}_j:=\{x\in\mathbb C:p_j(x)=0\}
$
is a finite set.
Consequently, $\bigcup_{j=1}^D  {\mathcal Z}_j \subset {\mathbb C}$ is finite, being the union of finite sets.
Hence, it suffices to choose
$
\bar\epsilon$ in 
${\mathbb R}\setminus  \bigcup_{j=1}^D  {\mathcal Z}_j$ to ensure that none of the $\mu_j, j\in [1,D],$ is an eigenvalue of $A-\bar \epsilon B$.
\\

$ii)\Rightarrow i)$ \ Is trivial. Indeed, for every $j\in [1,d]$ it is sufficient to choose $\epsilon_j = \bar \epsilon$.

\end{proof}

The following result is an easy corollary of Lemma \ref{techlemma2}, that can be proved by exploiting the proof of $i) \Rightarrow ii)$.

\begin{corollary}\label{cortechlemma2}
Given two $n\times n$ real matrices $A$ and $B$ and a family of (not necessarily distinct) complex numbers $\{\hat\lambda_2, \hat\lambda_3, \dots, \hat \lambda_N\}$,
suppose that 
$$\bigcap_{i=2}^N\sigma(A-\hat \lambda_i B) =\emptyset.$$
Then there exists a real number $\bar \epsilon \in {\mathbb R}$ such that $\sigma(A-\bar\epsilon B)\cap \left(\bigcup_{i=2}^N \sigma(A -\hat \lambda_i B)\right) =\emptyset$.
    \end{corollary}
   \begin{proof}
   Set
$$\bigcup_{i=2}^N\sigma(A-\hat \lambda_i B) =: \{\mu_1, \mu_2, \dots,  \mu_D\}.$$
If $\bigcap_{i=2}^N\sigma(A-\hat \lambda_i B) =\emptyset$,
then, for each $\mu_j, j\in [1,D],$   there exists 
$\hat \lambda_i, i\in [2,N]$, such that $\mu_j \not\in
\sigma(A-\hat \lambda_i B)$. This ensures that the polynomial
$
p_j(x):=\det(A-\mu_j I-xB)
$
is not identically zero, since $\det(A-\mu_j I- \hat \lambda_i B) \ne 0$,  and hence
$
{\mathcal Z}_j:=\{x\in\mathbb C:p_j(x)=0\}
$
is a finite set.
Consequently, $\bigcup_{j=1}^D  {\mathcal Z}_j \subset {\mathbb C}$ is finite, being the union of finite sets.
Hence, every
$
\bar\epsilon$ in 
${\mathbb R}\setminus  \bigcup_{j=1}^D  {\mathcal Z}_j$  ensures that none of the $\mu_j, j\in [1,D],$ is an eigenvalue of $A-\bar \epsilon B$.

\end{proof}

\begin{lemma}\label{lem.AB.obs.necessary}
 Given two $n\times n$ real matrices $A$ and $B$ and a family of (not necessarily distinct) complex numbers $\{\hat\lambda_2, \hat\lambda_3, \dots, \hat \lambda_N\}$,
if for every $\mu\in\bigcap_{i=2}^N \sigma(A-\hat \lambda_i B),$ there exists $\epsilon\in\mathbb{R}$ such that $\mu\notin\sigma(A-\epsilon B)$, then $(A,B)$ is an observable pair.
\end{lemma}
\begin{proof}
Suppose, by contradiction, that
$(A,B)$ is not observable, then there exists an eigenpair $(\mu,v)$ of $A$, $v\ne {\mathbb 0}$, such that $v\in\ker B$.
Consequently, 
\begin{align*}
(A-\hat \lambda_i B)v = Av = \mu v
\end{align*}
for every $i\in[2,N]$, and hence 
\begin{align*}
\mu\in\bigcap_{i=1}^N \sigma(A-\hat \lambda_i B).
\end{align*}
On the other hand,   for every $\epsilon\in\mathbb{R}$ it is also true that
\begin{align*}
(A-\epsilon B)v=\mu v
\end{align*}
This contradicts the hypothesis that $\mu\notin\sigma(A-\epsilon B)$. 

\end{proof}

\bibliographystyle{plain}
\bibliography{Refer192}

\end{document}